\documentclass{jfm}
\usepackage{graphicx}
\usepackage{epstopdf, epsfig}
\usepackage{xcolor}
\usepackage{amsmath}
\usepackage{mathrsfs}

\usepackage{ulem}
\newcommand\erase{\bgroup\markoverwith{\textcolor{red}{\rule[.5ex]{2pt}{0.8pt}}}\ULon}

\definecolor{cinnamon}{rgb}{0.82, 0.41, 0.12}
\def\v#1{{\boldsymbol{#1}}}  % vector
\def\m#1{{\mathcal{#1}}}

\shorttitle{Inertial focusing of spherical capsule in pulsatile channel flows}
\shortauthor{N. Takeishi, K. Ishimoto, N. Yokoyama and M. E. Rosti}

\title{Inertial focusing of spherical capsule in pulsatile channel flows}

\author{
  Naoki Takeishi\aff{1} \corresp{\email{takeishi.naoki.008@m.kyushu-u.ac.jp}},
  Kenta Ishimoto\aff{2},
  Naoto Yokoyama\aff{3} \and \\
  Marco Edoardo Rosti\aff{4}
}

\affiliation{
  \aff{1} Department of Mechanical Engineering, Kyushu University, 744 Motooka, Nishi-ku, Fukuoka 819-0395, Japan.
  \aff{2} Research Institute for Mathematical Sciences, Kyoto University, Kitashirakawa Oiwake-cho, Sakyo-ku, Kyoto 606-8502, Japan.
  \aff{3} Department of Mechanical Engineering, Tokyo Denki University, 5 Senju-Asahi, Adachi, Tokyo 120-8551, Japan.
  \aff{4} Complex Fluids and Flows Unit, Okinawa Institute of Science and Technology Graduate University, 1919-1 Tancha, Onna-son, Okinawa 904-0495, Japan.
}

\date{First submission DD MM. 2024; revised \today}
\begin{document}
\maketitle

\begin{abstract}
We present numerical analysis of the lateral movement of spherical capsule in the steady and pulsatile channel flow of a Newtonian fluid, for a wide range of oscillatory frequency. Each capsule membrane satisfying strain-hardening characteristic is simulated for different Reynolds numbers $Re$ and capillary numbers $Ca$.
%, the latter indicating the ratio of the fluid viscous force to the membrane elastic force.
Our numerical results showed that capsules with high $Ca$ exhibit axial focusing at finite $Re$ similarly to the inertialess case.
We observe that the speed of the axial focusing can be substantially accelerated by making the driving pressure gradient oscillating in time.
We also confirm the existence of an optimal frequency which maximises the speed of axial focusing, that remains the same found in the absence of inertia.
For relatively low $Ca$, on the other hand, the capsule exhibits off-centre focusing, resulting in various equilibrium radial positions depending on $Re$.
Our numerical results further clarifies the existence of a specific $Re$ for which the effect of the flow pulsation to the equilibrium radial position is maximum.
%Interestingly, such effective frequencies are in the same order of magnitude of that in axial focusing. Furthermore, the contribution of pulsatile flow to the equilibrium radial position is maximised at moderate Re, which allow the capsule to exhibit axial focusing in steady flow.
The roles of channel size on the lateral movements of the capsule are also addressed.
Throughout our analyses, we have quantified the radial position of the capsule in a tube based on an empirical expression. Given that the speed of inertial focusing can be controlled by the oscillatory frequency,
the results obtained here can be utilised for label-free cell alignment/sorting/separation techniques, e.g., for circulating tumor cells in cancer patients or precious hematopoietic cells such as colony-forming cells.
\end{abstract}


\section{Introduction}
In a pipe flow at a finite channel (or particle) Reynolds number $Re$ ($Re_p$), a rigid spherical particle exhibits migration perpendicular to the flow direction, originally reported by~\cite{Segre1962},
the so-called inertial focusing or tubular pinch effect,
where the particles equilibrate at a distance from the channel centreline as a consequence of the force balance between the shear-induced and wall-induced lift forces.
The phenomenon is of fundamental importance in microfluidic techniques such as label-free cell alignment, sorting, and separation techniques~\citep{Martel2014, Warkiani2016, Zhou2019}. 
%Applications of inertial microfluidics has been incfluencing a wide range of industries such as biology, chemistry, and biotechnology.
Although the techniques allow us to reduce the complexity and costs of clinical applications by using small amount of blood samples,
archetypal inertial focusing system requires steady laminar flow through long channel distances $L_f$, which can be estimated as $L_f = \pi \m{H}/(Re_p f_l)$, where $\m{H}$ is the dimension of the channel (or its hydraulic diameter) and $f_l$ is a non-dimensional lift coefficient~\citep{DiCarlo2009}.
So far, various kind of geometries have been proposed to achieve the required distance for inertial focusing in a compact space, e.g., sinusoidal, spiral, and hybrid channels~\citep{Bazaz2020}.
Without increasing $Re_p$, the recent experimental study by~\cite{Mutlu2018} achieved inertial focusing of $0.5$-$\mu$m-size particles ($Re_p \approx 0.005$) in short channels by using oscillatory channel flows.
Since the oscillatory flows allow a suspended particle to increase its total travel distance without net displacement along the flow direction,
utilizing oscillatory flow can be an alternative and practical strategy for inertial focusing in microfluidic devices.
Recently, \cite{Vishwanathan2021} experimentally investigated the effects of the Womersley number ($\alpha$) on inertial focusing in planar pulsatile flows,
and evaluated the lateral migration (or off-centre focusing) speed on a small and weakly inertial particle for different oscillatory frequencies.
They concluded that inertial focusing is achieved in only a fraction of the channel length ($1$ to $10$\%) compared to what would be required in a steady flow~\citep{Vishwanathan2021}.
%Inertial focusing in unbounded and wall-bounded steady flows are conceptually well understood and have been thoroughly reviewed~\citep{Stoecklein2019, Shi2020}.
\cite{Sun2009} performed two-dimensional (2D) simulations of a neutrally buoyant circular particle in oscillatory pressure-driven channel flows for $Re \geq 50$.
Their results indicated that lower oscillatory frequency makes the equilibrium position closer to the channel centerline while higher oscillatory frequency maintains the equilibrium positions similarly to the steady flow conditions.
However, it remains unknown whether the equilibrium position of deformable capsules under pulsatile channel flows can be formulated in the same context as that of rigid circular particle.

While a number of studies have analysed the off-centre focusing of rigid spherical particles under steady flow by a variety of approaches,
such as analytical calculations~\citep{Asmolov1999, Ho1974, Schonberg1989},
numerical simulations~\citep{Bazaz2020, Feng1994, Shao2008, Yang2005, Yu2004},
and experimental observations~\citep{DiCarlo2009, Karnis1966, Matas2004},
the inertial focusing of deformable particles such as biological cells,
consisting of an internal fluid enclosed by a thin membrane,
has not yet been fully described, especially under unsteady flows.
Due to their deformability,
the problem of inertial focusing of deformable particles is more complex than with rigid spherical particles, as originally reported by~\cite{Segre1962}.
%\cite{Jeffery1922} speculated that an ellipsoid may alter its orientation so that the viscous energy dissipation of the system becomes minimal.
%However, this is not true for soft particles with large deformation.
It is now well known that a deformable particle at low $Re$ migrates toward the channel axis under steady laminar flow~\citep{Karnis1963}.
Hereafter, we call this phenomenon as ``axial focusing''.
Recent numerical study showed that, in almost inertialess condition, the axial focusing of a deformable spherical capsule can be accelerated by the flow pulsation at a specific frequency~\citep{Takeishi2023}.
For finite $Re$ ($> 1$), however, it is still uncertain whether the flow pulsation can enhance the off-centre focusing or impede it (i.e., axial focusing).
Therefore, the primary objective of this study is to clarify whether a capsule lateral movement at finite $Re$ in a pulsatile channel flow can be altered by its deformability.
The second objective is to clarify whether the $Re$-dependent equilibrium radial position of a capsule in a channel or traveling time are controllable by oscillatory frequency.

At least for steady channel flows,
inertial focusing of deformable capsules including biological cells have been investigated in recent years both by means of experimental observations~\citep{Warkiani2016, Zhou2019} and numerical simulations~\citep{Raffiee2017, Schaaf2017, Takeishi2022}.
%So far, inertial focusing of rigid spherical particles have been well investigated, e.g., in~\cite{Martel2014, Morita2017, Nakagawa2015, Nakayama2019}.
%Studies of inertial focusing of biological cells have attracted particular attentions recently~\citep{Warkiani2016, Zhou2019}.
For instance, \cite{Hur2011} experimentally investigated the inertial focusing of various cell types (including red blood cells, leukocytes, and cancer cells such as a cervical carcinoma cell line, breast carcinoma cell line, and osteosarcoma cell line) with a cell-to-channel size ratio $0.1 \leq d_0/W \leq 0.8$,
using a rectangular channel with a high aspect ratio of $W/H \approx 0.5$,
where $d_0$, $W$ and $H$ are the cell equilibrium diameter, channel width, and height, respectively.
They showed that the cells can be separated according to their size and deformability~\citep{Hur2011}.
The experimental results can be qualitatively described using a spherical capsule~\citep{Kilimnik2011} or droplet model~\citep{Chen2014}.
In more recent experiments by~\cite{Hadikhani2018}, the authors investigated the effect of $Re$ ($1 < Re < 40$) and capillary number $Ca$ -- ratio between the fluid viscous force and the membrane elastic force -- ($0.1 < Ca < 1$) on the lateral equilibrium of bubbles in rectangular microchannels and different bubble-to-channel size ratios with $0.48 \leq d_0/W \leq 0.84$.
The equilibrium position of such soft particles results from the competition between $Re$ and $Ca$,
because high $Re$ induce the off-centre focusing,
while high $Ca$, i.e., high deformability, allows axial focusing.
However, notwithstanding these recent advancements, a comprehensive understanding of the effect on the inertial focusing of these two fundamental parameters has not been fully established yet.

Numerical analysis more clearly showed that the ``deformation-induced lift force'' becomes stronger as the particle deformation is increased~\citep{Raffiee2017, Schaaf2017}.
Although a number of numerical analyses regarding inertial focusing have been reported in recent years mostly for spherical particles~\citep{Bazaz2020, banerjee_rosti_kumar_brandt_russom_2021a},
the equilibrium positions of soft particles is still debated owing to the complexity of the phenomenon.
\cite{Kilimnik2011} showed that the equilibrium position in a cross section of rectangular microchannel with $d_0/H = 0.2$ shifts toward the wall as $Re$ increases from $1$ to $100$.
\cite{Schaaf2017} also performed numerical simulations of spherical capsules in a square channel for $0.1 \leq d_0/H \leq 0.4$ and $5 \leq Re \leq 100$ without viscosity contrast,
and showed that the equilibrium position was nearly independent of $Re$.
In a more recent numerical analysis by~\cite{Alghalibi2019},
simulations of a spherical hyperelastic particle in a circular channel with $d_0/D = 0.2$ were performed with $100 \leq Re \leq 400$ and Weber number ($We$) with $0.125 \leq We \leq 4.0$,
the latter of which is the ratio of the inertial effect to the elastic effect acting on the particles.
Their numerical results showed that regardless of $Re$,
the final equilibrium position of a deformable particle is the centreline,
and harder particles (i.e., with lower $We$) tended to rapidly migrate toward the channel centre~\citep{Alghalibi2019}.
The behaviour of a capsule subjected to pulsatile channel flow was addressed in the pioneering work by~\cite{Maestre2019},
where the migration velocity during axial focusing was investigated in $O(Re) \leq 10^{-2}$ and $d_0/D = 0.5$ for $Ca = 0.075$--$1.2$.
Despite these efforts,
the inertial focusing of capsules subjected to pulsatile flow at finite inertia cannot be estimated based on these achievements.

Aiming for the precise description of the inertial focusing of spherical capsules in pulsatile channel flows, 
we thus perform numerical simulations of individual capsules with a major diameter of $d_0 = 2a_0 = 8$ $\mu$m in a cylindrical microchannel with $D = 2R = 20$--$50$ $\mu$m (i.e., $R/a_0 = 2.5$--$6.25$) for a wide range of oscillatory frequency.
Each capsule membrane, following the Skalak constitutive (SK) law~\citep{Skalak1973}, is simulated for different $Re$, $Ca$, and size ratio $R/a_0$
%The membrane of the capsule follows the Skalak constitutive (SK) law~\citep{Skalak1973}.
Since this problem requires heavy computational resources,
we resort to GPU computing,
using the lattice-Boltzmann method (LBM) for the inner and outer fluids and the finite element method (FEM) to describe the deformation of the capsule membrane.
This model has been successfully applied in the past for the analysis of the capsule flow in circular microchannels~\citep{Takeishi2022, Takeishi2023}.
The remainder of this paper is organised as follows.
Section $2$ gives the problem statement and numerical methods,
Section $3$ presents the numerical results for single spherical capsule. Finally, a summary of the main conclusions is reported in Section $4$.
A description of numerical verifications is presented in the Appendix.

\section{Problem statement}
\subsection{Flow and capsule models and setup}
We consider the motion of an initially spherical capsule with diameter $d_0$ (= 2$a_0$ = 8 $\mu$m) flowing in a circular channel diameter $D$ (= 2$R$ = 20--50 $\mu$m),
with a resolution of 32 fluid lattices per capsule diameter $d_0$.
The channel length is set to be 20$a_0$, following previous numerical study~\citep{Takeishi2022}.
Although we have investigated in the past the effect of the channel length $L$ and the mesh resolutions on the trajectory of the capsule centroid (see Fig.~7 in~\cite{Takeishi2023}),
we further assess the effect of this length on the lateral movement of a capsule in Appendix~\S\ref{appA1} (figure~\ref{fig:verification}$a$).
\begin{figure}%Figure 1
  \centering
  \includegraphics[height=5cm]{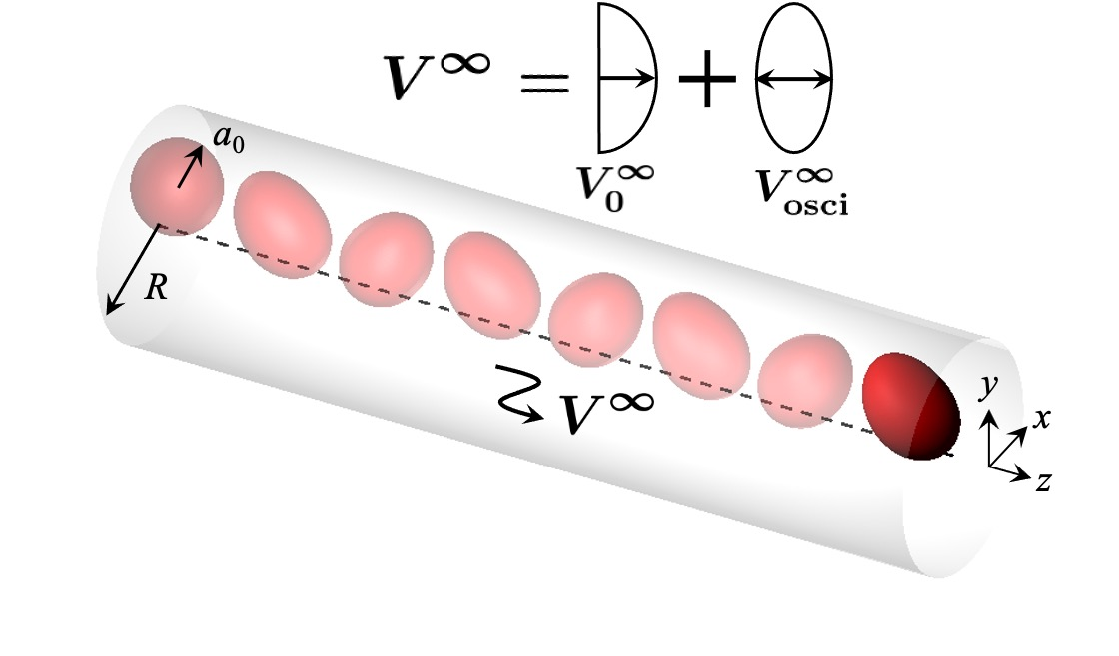}
  \caption{
         Visualisation of a spherical capsule with radius $a_0$ in a channel with radius of $R$ ($R/a_0$ = 2.5) under a pulsatile flow with velocity $V^\infty$,
         which can be decomposed into the steady parabolic flow $V_0^\infty$ and the oscillatory flow $V_\mathrm{osci}^\infty$ in the absence of any capsule.
         The capsule, initially placed at off-centre radial position $r^\ast_\mathrm{c0}= r_\mathrm{c0}/R = 0.4$, travels in the radial direction.
         In the figure, the lengths travelled by the capsule in the flow ($z$) direction is not to scale for illustrative purpose.
         Hereafter, the same modification will be applied for visualisation.
         %, where a material points at the initial spherical top node point is represented by green dots.
  }
  \label{fig:snapshot}
\end{figure}

The capsule consists of a Newtonian fluid enclosed by a thin elastic membrane, sketched in figure~\ref{fig:snapshot}.
%where the length scales of the capsule centroid in the flow direction ($z$ direction) are modified for an illustrate purpose.
%Hereafter, the same modification is applied for visualisation.
%, where a material points at the initial spherical top node point is represented by green dots.
The membrane is modeled as an isotropic and hyperelastic material following the SK law~\citep{Skalak1973}, in which the strain energy $w_\mathrm{SK}$ and principal tensions in the membrane $\tau_1$ and $\tau_2$ (with $\tau_1 \geq \tau_2$) are given by
\begin{equation}
  \frac{w_\mathrm{SK}}{G_s} = \frac{1}{4} \left( I_1^2 + 2I_1 - 2I_2 + C I_2^2\right),
  \label{SK}
\end{equation}
and
\begin{equation}
  \frac{\tau_i}{G_s} = \frac{\eta_i}{\eta_j} \left[ \eta_i^2 - 1 + C \left( \eta_i^2 \eta_j^2 - 1 \right) \right], \quad \text{for} \ (i, j) = (1, 2) \ \text{or} \ (2, 1).
  \label{SK2}
\end{equation}
Here, $w_\mathrm{SK}$ is the strain energy density function,
$G_s$ is the membrane shear elastic modulus,
$C$ is a coefficient representing the area incompressibility,
$I_1$ ($= \eta_1^2 + \eta_2^2 - 2$) and $I_2$ ($= \eta_1^2\eta_2^2 - 1$) are the invariants of the strain tensor,
with $\eta_1$ and $\eta_2$ being the principal extension ratios.
In the SK law~\eqref{SK},
the area dilation modulus is $K_s = G_s (1 + 2C)$.
In this study, we set $C = 10^2$~\citep{BarthesBiesel2002},
which describes an almost incompressible membrane.
Bending resistance is also considered~\citep{Li2005},
with a bending modulus $k_b = 5.0 \times 10^{-19}$ J~\citep{Puig-de-Morales-Marinkovic2007}.
These values have been shown to successfully reproduce the deformation of red blood cells in shear flow~\citep{Takeishi2014, Takeishi2019} and the thickness of cell-depleted peripheral layer in circular channels (see Figure~A.1 in~\cite{Takeishi2014}).
Neglecting inertial effects on the membrane deformation,
the static local equilibrium equation of the membrane is given by
\begin{equation}
  \nabla_s \cdot \v{\tau} + \v{q} = \v{0},
  \label{StrongForm}
\end{equation}
where $\nabla_s (= \left( \v{I} - \v{n} \v{n} \right) \cdot \nabla)$ is the surface gradient operator,
$\v{n}$ is the unit normal outward vector in the deformed state,
$\v{q}$ is the load on the membrane,
and $\v{\tau}$ is the in-plane elastic tension that is obtained using the SK law (equation~\ref{SK}).

The fluids are modeled with the incompressible Navier--Stokes equations for the fluid velocity $\v{v}$:
\begin{align}
	\rho \left( \frac{\partial \v{v}}{\partial t} + \v{v} \cdot \nabla \v{v} \right)
	&= \nabla \cdot \v{\sigma}^f  + \rho \v{f}, \\
	\nabla \cdot \v{v} &= 0,
\end{align}
with 
\begin{align}
\v{\sigma}^f = -p\v{I} + \mu \left( \nabla \v{v} + \nabla \v{v}^T \right),
\end{align}
where $\v{\sigma}^f$ is the total stress tensor of the flow,
$p$ is the pressure, 
$\rho$ is the fluid density,
$\v{f}$ is the body force,
and $\mu$ is the viscosity of the liquid,
expressed using a volume fraction of the inner fluid $\m{I}$ (0 $\leq \m{I} \leq$ 1) as:
\begin{align}
        \mu = \left\{ 1 + \left( \lambda - 1 \right) \m{I} \right\} \mu_0,
\end{align}
where $\lambda$ (= $\mu_1/\mu_0$) is the viscosity ratio, 
$\mu_0$ is the external fluid viscosity, and $\mu_1$ is the internal fluid viscosity.
No density contrast is considered; that is, the ratio of densities between the external and internal fluid is assumed to be one.

The dynamic condition coupling the different phases requires the load $\v{q}$ to be equal to the traction jump $\left( \v{\sigma}^f_\mathrm{out} - \v{\sigma}^f_\mathrm{in} \right)$ across the membrane:
\begin{align}
	\v{q} = \left( \v{\sigma}^f_\mathrm{out} - \v{\sigma}^f_\mathrm{in} \right) \cdot \v{n},
\end{align}
where the subscripts `out' and `in' represent the outer and internal regions of the capsule, respectively.

The flow in the channel is sustained by a uniform pressure gradient $\partial p_0/\partial z (= \partial_z p_0)$, 
which can be related to the maximum fluid velocity in the channel by $\partial_z p_0 = -4 \mu_0 V_\mathrm{max}^\infty/R^2$.
The pulsation is given by a superimposed sinusoidal function,
such that the total pressure gradient is
\begin{equation}
  \partial_z p (t) = \partial_z p_0 + \partial_z p_a \sin{\left( 2 \pi f t \right)}.
  \label{p_grad}
\end{equation}

%The problem is characterised by the Reynolds number $Re$ and by the capillary number $Ca$, defined as
The problem is governed by six main non-dimensional numbers, including $i$) the Reynolds number $Re$ and $i\hspace{-1.0pt}i$) the capillary number $Ca$ defined as:
\begin{align}
  &Re = \frac{\rho D V_{\mathrm{max}}^{\infty}}{\mu_0}, \\
  &Ca = \frac{\mu_0 \dot{\gamma}_\mathrm{m} a_0}{G_s} = \frac{\mu_0 V_{\mathrm{max}}^{\infty}}{G_s} \frac{a_0}{4 R},
\end{align}
where $V_{\mathrm{max}}^{\infty}$ ($= 2V_\mathrm{m}^{\infty})$ is the maximum fluid velocity in the absence of any cells,
$V_\mathrm{m}^{\infty}$ is the mean fluid velocity,
and $\dot{\gamma}_\mathrm{m}$ ($= V_\mathrm{m}^{\infty}/D)$ is the mean shear rate.
Note that, increasing $Re$ under constant $Ca$ corresponds to increasing $G_s$, namely, a harder capsule.
Furthermore, we have $i\hspace{-1.0pt}i\hspace{-1.0pt}i)$ the viscosity ratio $\lambda$, $i\hspace{-1.0pt}v$) the size ratio $R/a_0$, $v$) the non-dimensional pulsation frequency $f^\ast = f/\dot{\gamma}_\mathrm{m}$, and $v\hspace{-1.0pt}i$) the non-dimensional pulsation amplitude $\partial_z p_a^\ast = \partial_z p_a/\partial_z p_0$. Considered the focus of this study, we decide to primarily investigate the effect of $Re$, $R/a_0$, and $f^\ast$.
Representative rigid and largely deformable capsules are considered with $Ca = 0.05$ and $Ca = 1.2$, respectively.

When presenting the results, we will initially focus on the analysis of lateral movements of the capsule in effectively inertialess condition ($Re = 0.2$) for $R/a_0 = 2.5$, and later consider variations of the size ratio $R/a_0$, viscosity ratio $\lambda$, Reynolds number $Re$ ($> 1$), and $Ca$.
We confirmed that the flow at $Re = 0.2$ well approximates an almost inertialess flow for single-~\citep{Takeishi2023} and multi-cellular flow~\citep{Takeishi2019}.
Unless otherwise specified, we show the results obtained with $\partial_z p_a^\ast = 2$ and $\lambda = 1$.

\subsection{Numerical simulation}
The governing equations for the fluid are discretised by the LBM based on the D3Q19 model~\citep{Chen1998}.
We track the Lagrangian points of the membrane material points $\v{x}_m (\v{X}_m,t)$ over time,
where $\v{X}_m$ is a material point on the membrane in the reference state.
Based on the virtual work principle,
the above strong-form equation (\ref{StrongForm}) can be rewritten in weak form as 
\begin{equation}
  \int_S \v{\hat{u}} \cdot \v{q} dS = \int_S \v{\hat{\epsilon}} : \v{\tau} dS,
  \label{WeakForm}
\end{equation}
where $S$ is the surface area of the capsule membrane, and $\v{\hat{u}}$ and $\v{\hat{\epsilon}} = ( \nabla_s \v{\hat{u}} + \nabla_s \v{\hat{u}}^T )\big/2$ are the virtual displacement and virtual strain, respectively.
The FEM is used to solve equation (\ref{WeakForm}) and obtain the load $\v{q}$ acting on the membrane~\citep{Walter2010}.
The velocity at the membrane node is obtained by interpolating the velocities at the fluid node using the immersed boundary method~\citep{Peskin2002}.
The membrane node is updated by Lagrangian tracking with the no-slip condition.
The explicit fourth-order Runge--Kutta method is used for the time integration. 
The volume-of-fluid method~\citep{Yokoi2007} and front-tracking method~\citep{Unverdi1992} are employed to update the viscosity in the fluid lattices.
A volume constraint is implemented to counteract the accumulation of small errors in the volume of the individual cells~\citep{Freund2007}:
in our simulation, the relative volume error is always maintained lower than $1.0 \times 10^{-3}$\%,
as tested and validated in our previous study of cell flow in circular channels~\citep{Takeishi2016}.
All procedures were fully implemented on a GPU to accelerate the numerical simulation.
More precise explanations for numerical simulations including membrane mechanics are provided in our previous works~\citep[see also][]{Takeishi2019, Takeishi2022}.

Periodic boundary conditions are imposed in the flow direction ($z$-direction).
No-slip conditions are employed for the walls (radial direction).
We set the mesh size of the LBM for the fluid solution to $250$ nm,
and that of the finite elements describing the membrane to approximately $250$ nm (an unstructured mesh with $5120$ elements was used for the FEM).
This resolution was shown to successfully represent single- and multi-cellular dynamics~\citep{Takeishi2019, Takeishi2022}.

\subsection{Analysis of capsule deformation}
Later, we investigate the in-plane principal tension $T_i$ (with $T_1 \geq T_2$) and the isotropic tension $T_\mathrm{iso}$ in the membrane of the capsule.
In the case of a two-dimensional isotropic elastic membrane,
the isotropic membrane tension can be calculated by $T_\mathrm{iso} = (T_1 + T_2)/2$ for the deformed capsule.
The averaged value of $T_\mathrm{iso}$ is then calculated as
\begin{equation}
  \langle T_\mathrm{iso} \rangle = \frac{1}{S\m{T}} \int_\m{T} \int_S T_\mathrm{iso} (\v{x}_m, t) dS dt,
  \label{Tiso}
\end{equation}
where $\m{T}$ is the period of the capsule motion.
Hereafter, $\langle \cdot \rangle$ denotes a spatial-temporal average.
Time average starts after the trajectory has finished the initial transient dynamics,
which differs for each case.
For instance, at finite $Re$ conditions, a quasi-steady state is usually attained around the non-dimensional time of $\dot{\gamma}_\mathrm{m} t = 200$,
and we start accumulating the statistics from $\dot{\gamma}_\mathrm{m} t \geq 400$ to fully cancel the influence of the initial conditions. 
\begin{figure}%Figure 2
  \centering
  \includegraphics[height=6cm]{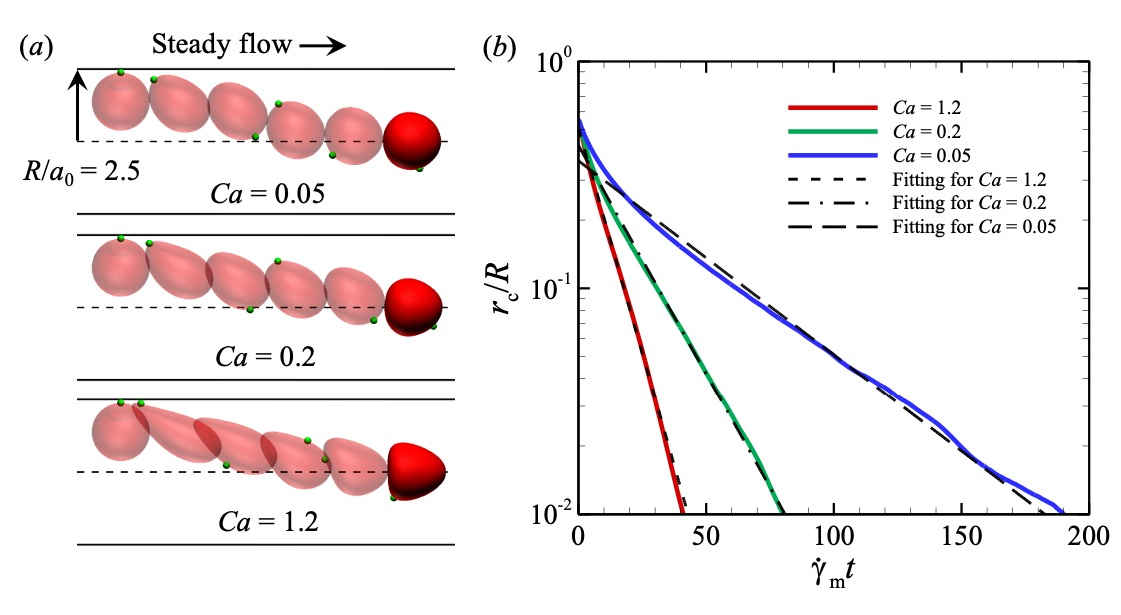}
  \caption{
         ($a$) Side views of the capsule during its axial focusing under steady flow for $Ca = 0.05$ (top), $Ca = 0.2$ (middle), and $Ca = 1.2$ (bottom).
         The capsule is initially placed at $r^\ast_\mathrm{c0} = 0.55$.
         The coloured dot on the membrane is shown to measure the membrane rotation.
         ($b$) Time histories of the radial position of these capsule centroids $r_\mathrm{c}/R$.
         The dashed lines are the curves given by $r_\mathrm{c}^\ast = C_2 \exp{(-C_1 t^\ast)}$,
         where $r_\mathrm{c}^\ast$ ($= r_\mathrm{c}/R$) is the non-dimensional capsule centroid,
         $t^\ast$ ($= \dot\gamma_\mathrm{m}t$) is the non-dimensional time,
         and $C_1$ and $C_2$ are the coefficients found by a least-squares fitting to the plot.
 	  The results in the figure are obtained for $Re= 0.2$, $R/a_0 = 2.5$, and $\lambda = 1$.
  }
  \label{fig:fitting}
\end{figure}

\section{Results}
\subsection{\label{S1}Axial focusing of the capsule under steady channel flow ($Re < 1$)}
We first investigate the axial focusing of a capsule under steady flow,
which can be assumed to be effectively inertialess ($Re = 0.2$).
Figure~\ref{fig:fitting}($a$) shows side views of the capsule during its axial focusing in channel of size $R/a_0 = 2.5$ for different $Ca$ ($= 0.05$, $0.2$, and $1.2$).
The capsule, initially placed at $r^\ast_\mathrm{c0} = r_\mathrm{c0}/R = 0.55$,
migrates after the flow onsets towards the channel centreline (i.e., capsule centroid is $r_\mathrm{c} = 0$) while deforming,
finally reaching its equilibrium position at the centreline
where it achieves an axial-symmetric shape.
Although the magnitude of deformation during axial focusing depends on $Ca$,
these process is commonly observed for every $Ca$.
The time history of the radial position of the capsule centroid $r_\mathrm{c}$ is shown in figure~\ref{fig:fitting}($b$).
The results clearly show that the speed of axial focusing grows with $Ca$.
Interestingly, all trajectories are well fitted by the following empirical expression:
\begin{equation}
  r_\mathrm{c}^\ast = C_2 \exp{(-C_1 t^\ast)},
  \label{fitting}
\end{equation}
where $t^\ast$ ($= \dot\gamma_\mathrm{m}t$) is the non-dimensional time,
and $C_1$ ($> 0$) and $C_2$ are two coefficients that can be found by a least-squares fitting to the plot.
Fitting are performed using data between the initial ($r^\ast_\mathrm{c0} = 0.55$) and final state ($\Delta x_\mathrm{LBM}/R \leq 0.01$ for $R/a_0 = 2.5$),
defined as the time when the capsule is within one mesh size ($\Delta x_\mathrm{LBM}$) from the channel axis.

Performing time differentiation of equation~\eqref{fitting},
the non-dimensional velocity of the capsule centroid $\dot{r}_c^\ast$ can be estimated as:
\begin{equation}
  \dot{r}_\mathrm{c}^\ast = -C_1 r_\mathrm{c}^\ast.
  \label{lift}
\end{equation}
This linear relation \eqref{lift} may be understood by a shear-induced lift force propotional to the local shear strength. A more detailed description of the relationship between the coefficient $C_1$ and the lift force on the capsule are provided in Appendix \S\ref{appA_Lift}.

Figure~\ref{fig:alpha}($a$) shows the coefficient $C_1$ as a function of $Ca$.
As expected from figure~\ref{fig:fitting}($b$),
the value of $C_1$ increases with $Ca$.
Since the capsule deformability is also affected by the viscosity ratio $\lambda$,
its influence on $C_1$ is also investigated in figure~\ref{fig:alpha}($b$).
At a fixed $Ca$ ($= 1.2$),
the value of $C_1$ decreases with $\lambda$.
%We have thus numerically proven the existence of the coefficient $C_1$ scaling the lift force acting on the capsule membrane. %\nt{XXX $<$ WHY? XXX}
\begin{figure}%Figure 3
  \centering
  \includegraphics[height=4.5cm]{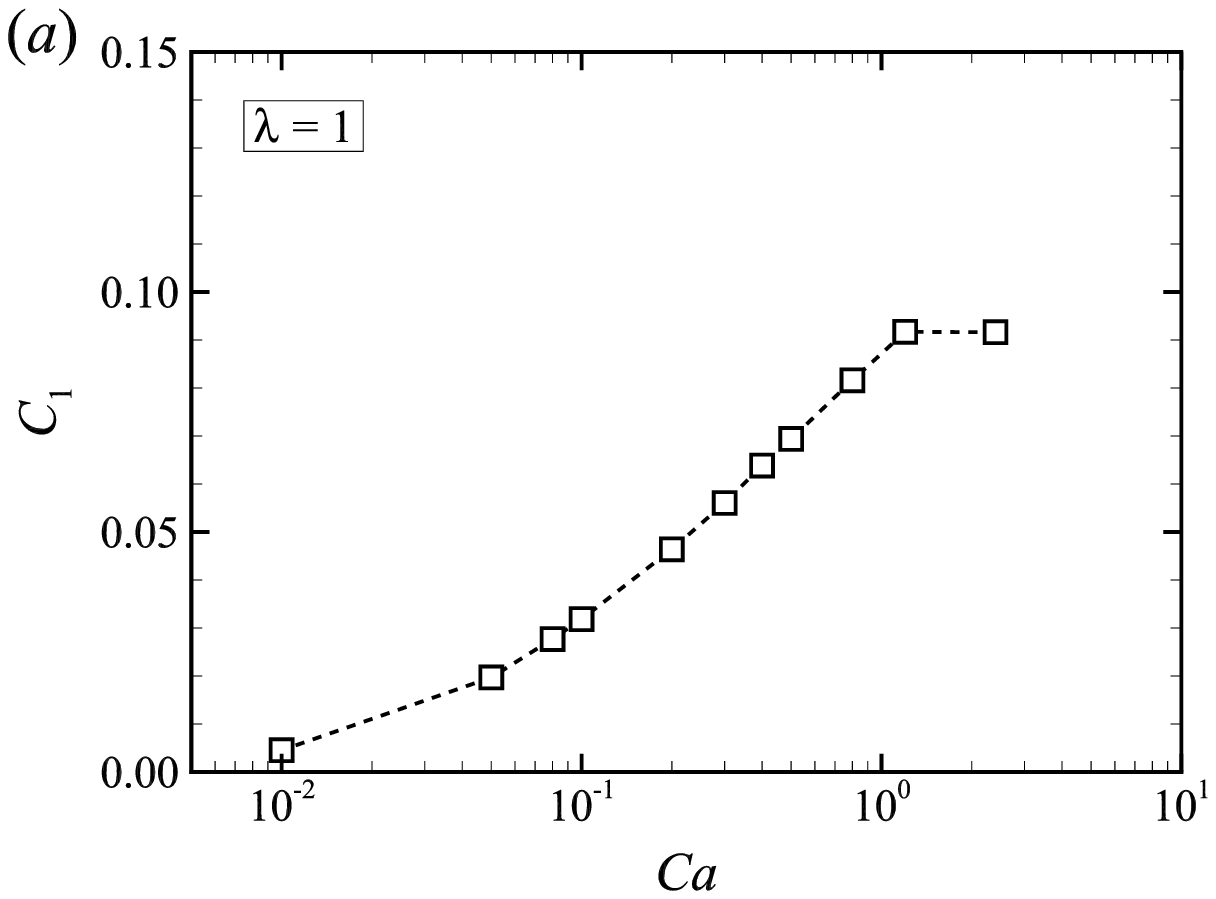}
  \includegraphics[height=4.5cm]{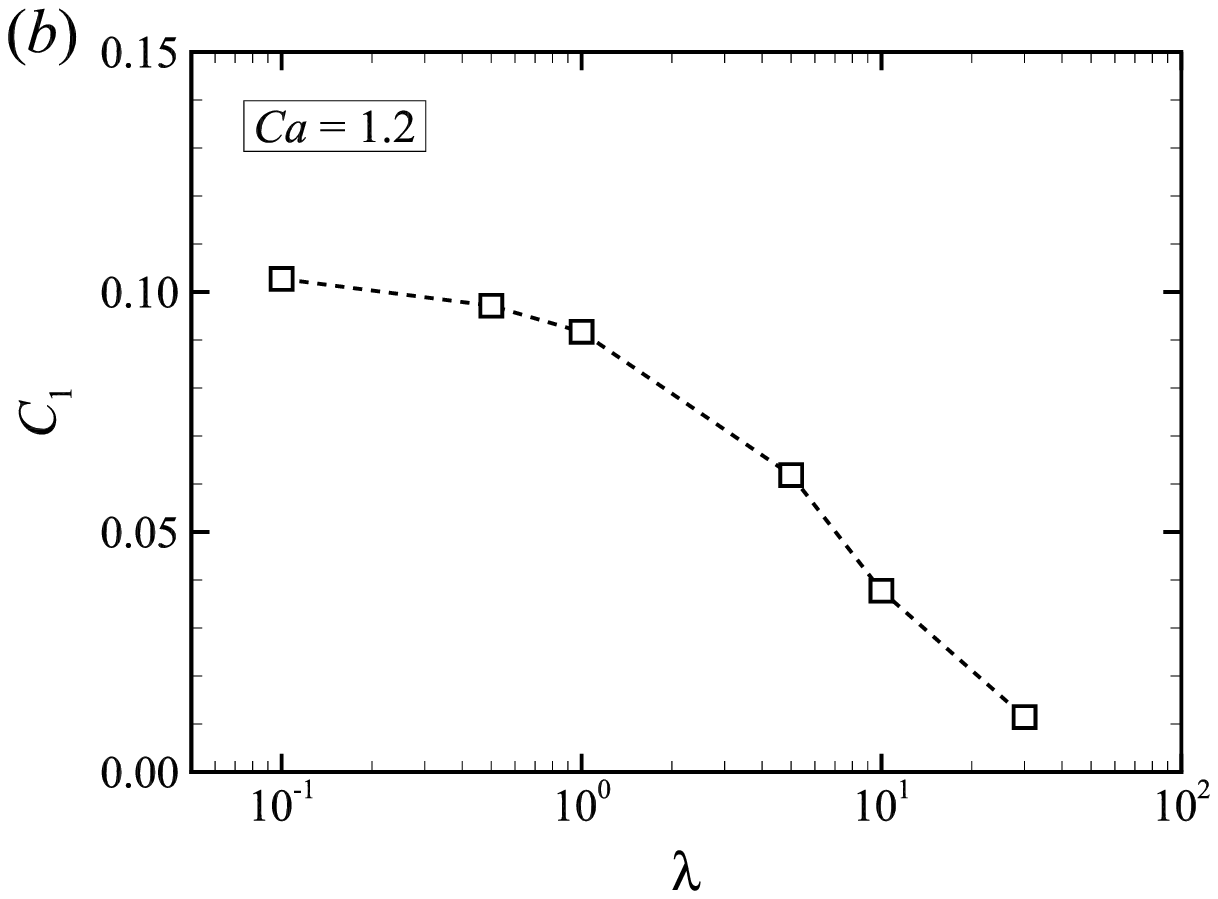}
  \caption{
         The coefficient $C_1$ ($a$) as a function of $Ca$ for $\lambda = 1$,
         and ($b$) as a function of $\lambda$ for $Ca = 1.2$.
 	  The results are obtained with $Re = 0.2$, $R/a_0 = 2.5$, and $r^\ast_\mathrm{c0} = 0.55$.
  }
  \label{fig:alpha}
\end{figure}

\begin{figure}%Figure 4
  \centering
  \includegraphics[height=8cm]{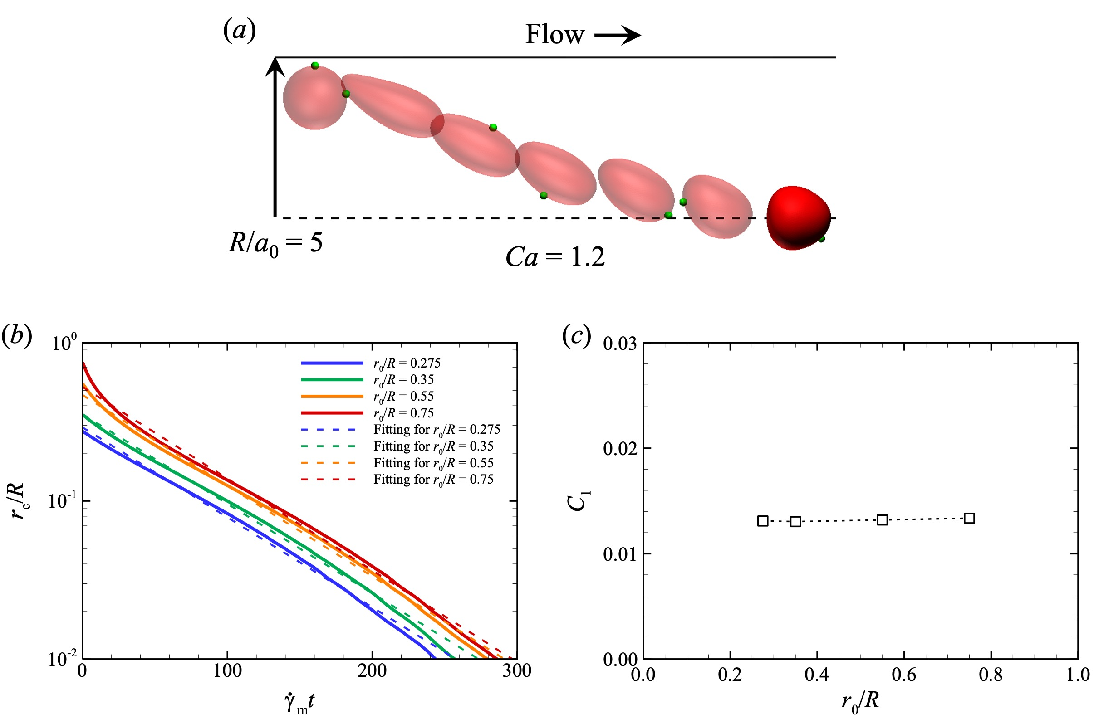}
  \caption{
         ($a$) Side views of a capsule with $Ca = 1.2$ during its axial focusing for $R/a_0 = 5$,
         where the capsule is initially placed at $r^\ast_\mathrm{c0} = 0.75$.
         ($b$) Time histories of the radial position of the capsule centroids $r_\mathrm{c}/R$ for different initial positions $r^\ast_\mathrm{c0}$.
         ($c$) The coefficient $C_1$ as a function of the initial position $r^\ast_\mathrm{c0}$.
 	  The results are obtained with $\lambda = 1$.
  }
  \label{fig:alpha_large_channel}
\end{figure}

To further proof that $C_1$ is independent of the initial radial position of the capsule centroid,
additional numerical simulations are performed with a larger channel ($R/a_0 = 5$) for different $r^\ast_\mathrm{c0}$.
Note that a case with larger channel for constant $Re$ denotes smaller $V_\mathrm{max}^\infty$, resulting in smaller $G_s$ (i.e., softer capsule) for constant $Ca$.
Figure~\ref{fig:alpha_large_channel}($a$) is one of the additional runs at $Ca = 0.2$,
where the capsule is initially placed at $r^\ast_\mathrm{c0} = 0.75$.
Figure~\ref{fig:alpha_large_channel}($b$) is the time history of the radial position of the capsule centroid $r_\mathrm{c}$ for different initial positions $r^\ast_\mathrm{c0}$.
We observe that the exponential fitting is still applicable for these runs, with the coefficient $C_1$ reported in figure~\ref{fig:alpha_large_channel}($c$).
These results provide a confirmation that $C_1$ is indeed independent of the initial radial position $r^\ast_\mathrm{c0}/R$.
Furthermore, the fitting provided in equation~\eqref{fitting} is applicable even for a different constitutive law.
Discussion of these results for capsule described by the neo-Hookean model,
which features strain-softening,
is reported in Appendix~\S\ref{appA2} (see also figure~\ref{fig:fitting_NH}).

\subsection{Capsule behaviour under pulsatile channel flow}
\begin{figure}%figure 5
  \centering
  \includegraphics[height=13cm]{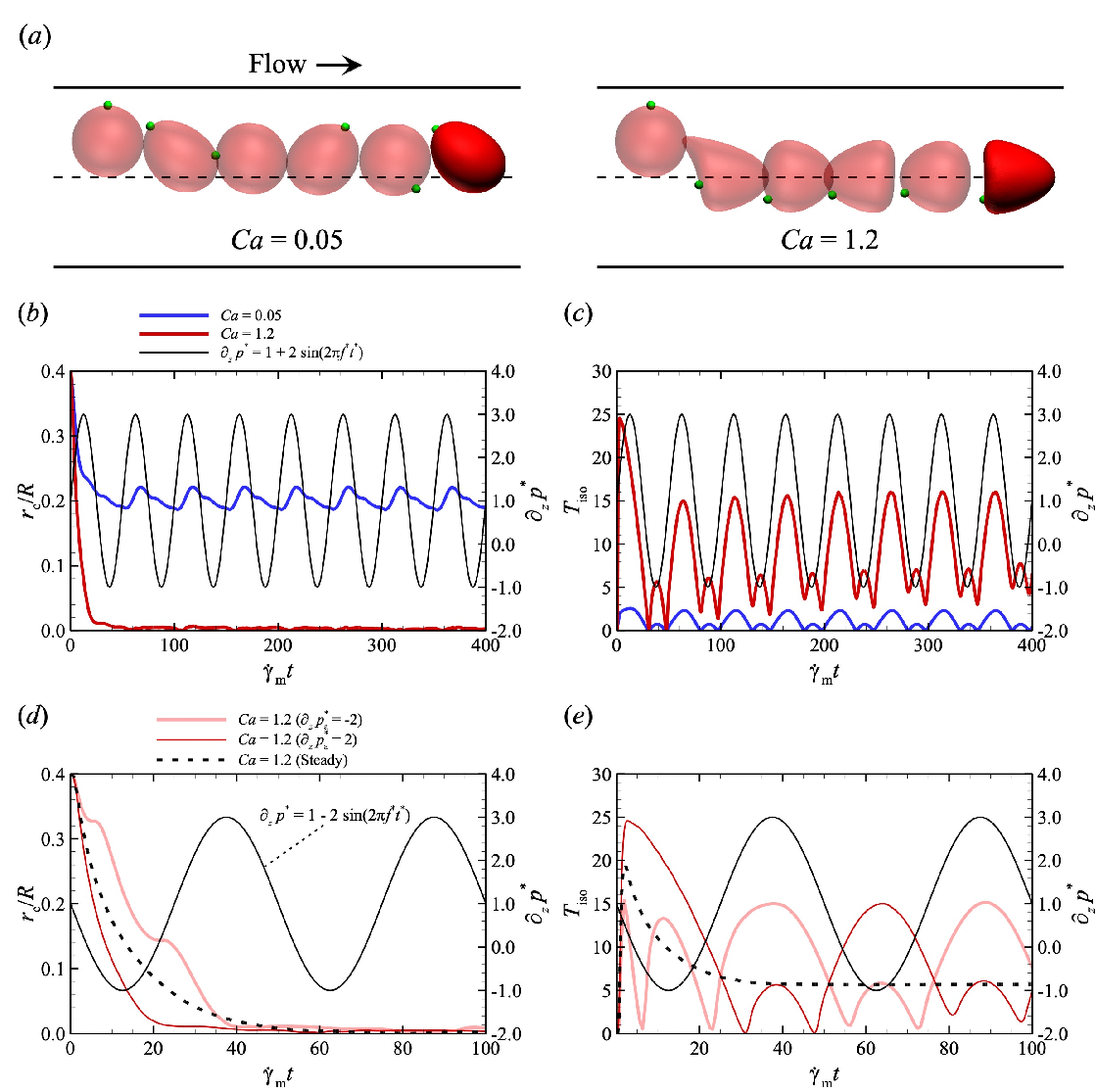}
  \caption{
         ($a$) Side views of the capsule during its migration at each time at $f^\ast = 0.02$ for $Ca = 0.05$ (left; see the supplementary movie 1, available at https://doi.org/xxx/jfm.2024.yyy) and $Ca = 1.2$ (right; see the supplementary movie 2).
         ($b$ and $c$) Time histories of ($b$) the radial position of these capsule centroids $r_\mathrm{c}/R$ and ($c$) isotropic tensions $T_\mathrm{iso}$, respectively.
         In panels ($a$--$c$), the results are obtained with $\partial_z p_a^\ast = 2$.
         ($d$ and $e$) Time histories of $r_\mathrm{c}/R$ and $T_\mathrm{iso}$ for $\partial_z p_a^\ast = -2$, respectively,
         where those in steady flow are also superposed.
         In panels ($b$--$e$), non-dimensional pressure gradient $\partial_z p^\ast$ is also displayed on right axis.
 	  The results are obtained with $Re = 10$, $R/a_0 = 2.5$, and $r^\ast_\mathrm{c0} = 0.4$.
  }
  \label{fig:re10f002}
\end{figure}

Next, we investigate inertial focusing of capsules at finite $Re$,
and investigate whether the equilibrium radial position of the capsule can be altered by pulsations of the flow.
%It is expected that the equilibrium radial position of capsules results from the competition between $Re$ and $Ca$, because at high $Re$, the flow pushes the particles towards the wall, while at high $Ca$, i.e. high deformability, particles can move towards the channel centre.
Two representative behaviours of the capsule at low $Ca$ ($= 0.05$) and high $Ca$ ($= 1.2$) are shown in figure~\ref{fig:re10f002}($a$),
which are obtained with $f^\ast = 0.02$ and $Re = 10$.
The simulations are started from a off-centre radial position $r^\ast_\mathrm{c0}$.
Hereafter, we consider the viscosity ratio $\lambda = 1$ for simplicity.
At the end of the migration, the least deformable capsule ($Ca = 0.05$) exhibits an ellipsoidal shape with an off-centred position (figure~\ref{fig:re10f002}$a$, left), while the most deformable one ($Ca = 1.2$) exhibits the typical parachute shape at the channel centreline (figure~\ref{fig:re10f002}$a$, right).
Detailed trajectories of these capsule centroids $r_\mathrm{c}/R$ are shown in figure~\ref{fig:re10f002}($b$),
where the non-dimensional oscillatory pressure gradient $\partial_z p^\ast (t^\ast)$ ($= 1 + 2 \sin{(2 \pi f^\ast t^\ast)}$) is also displayed.
The least deformable capsule ($Ca = 0.05$) fluctuates around the off-centre position $r_\mathrm{c}/R$ ($\approx 0.2$),
and the waveform of $r_\mathrm{c}/R$ lags behind $\partial_z p^\ast (t^\ast)$.
The capsule with large $Ca$ ($= 1.2$), on the other hand, immediately exhibits axial focusing, reaching the centerline within one flow period (figure~\ref{fig:re10f002}$b$).
Therefore, axial and off-centre focusing strongly depend on $Ca$.

Figure~\ref{fig:re10f002}($c$) is the time history of the isotropic tension $T_\mathrm{iso}$.
The major waveforms of $T_\mathrm{iso}$ are synchronised with $\partial_z p^\ast$ in both $Ca = 0.05$ and $Ca = 1.2$, thus indicating that the membrane tension spontaneously responds to the background fluid flow.
The Taylor parameter, a classical index of deformation, is described in Appendix \S\ref{appA3} (see figure~\ref{fig:D12_time}).

To clarify whether fast axial focusing depends on the phase of oscillation or not,
an antiphase pulsation (i.e., $\partial_z p_a^\ast = -2$) is given by $\partial_z p^\ast (t^\ast) = 1 - 2 \sin{(2 \pi f^\ast t^\ast)}$.
Time histories of the capsule centroid $r_\mathrm{c}/R$ and membrane tension $T_\mathrm{iso}$ under such condition are shown in figures~\ref{fig:re10f002}($d$) and \ref{fig:re10f002}($e$),
where the case at the same $Ca = 1.2$ from figures~\ref{fig:re10f002}($b$) and \ref{fig:re10f002}($c$) are also superposed for comparison, together with the solution for steady flow.
Here, we define the focusing times $T$ and $T_\mathrm{st}$ needed by the capsule centroid to reach the centreline (within a one fluid mesh corresponding to $\sim 6\%$ of its radius to account for the oscillations in the capsule trajectory) under pulsatile and steady flows, respectively.
Although the focusing time is decreased almost by $50$\% in prograde pulsation ($\partial_z p_a^\ast = 2$) compared to that in the steady flow,
the time in antiphase pulsation is decreased only by $1$\%.
Such small acceleration in antiphase pulsation comes from relatively small deformation in early periods (figure~\ref{fig:re10f002}$e$).
We now understand that fast axial focusing relies on the large membrane tension after flow onset,
and our numerical results exhibit the even faster axial focusing due to the pulsation of the flow.
%\erase{Although the capsule subject to antiphase pulsation takes longer time to complete the axial focusing than with prograde pulsation (i.e., $\partial_z p_a^\ast = 2$),
%the focusing speed is still slightly faster than that in steady flow (figure~\ref{fig:re10f002}$d$).
%This is due to the tension on the deformed membrane being greater than in steady flow, with an amplitude $T_\mathrm{iso}$ equal to those in prograde pulsation except for early periods (figure~\ref{fig:re10f002}$e$).}
%\erase{Figures~\ref{fig:re10f002}($d$) and \ref{fig:re10f002}($e$) suggest that the pulsation itself can enhance the axial focusing.}
%Hereafter, we show the results obtained with the prograde pulsation (i.e., $\partial_z p_a^\ast = 2$).

\begin{figure}%figure 6
  \centering
  \includegraphics[height=4.5cm]{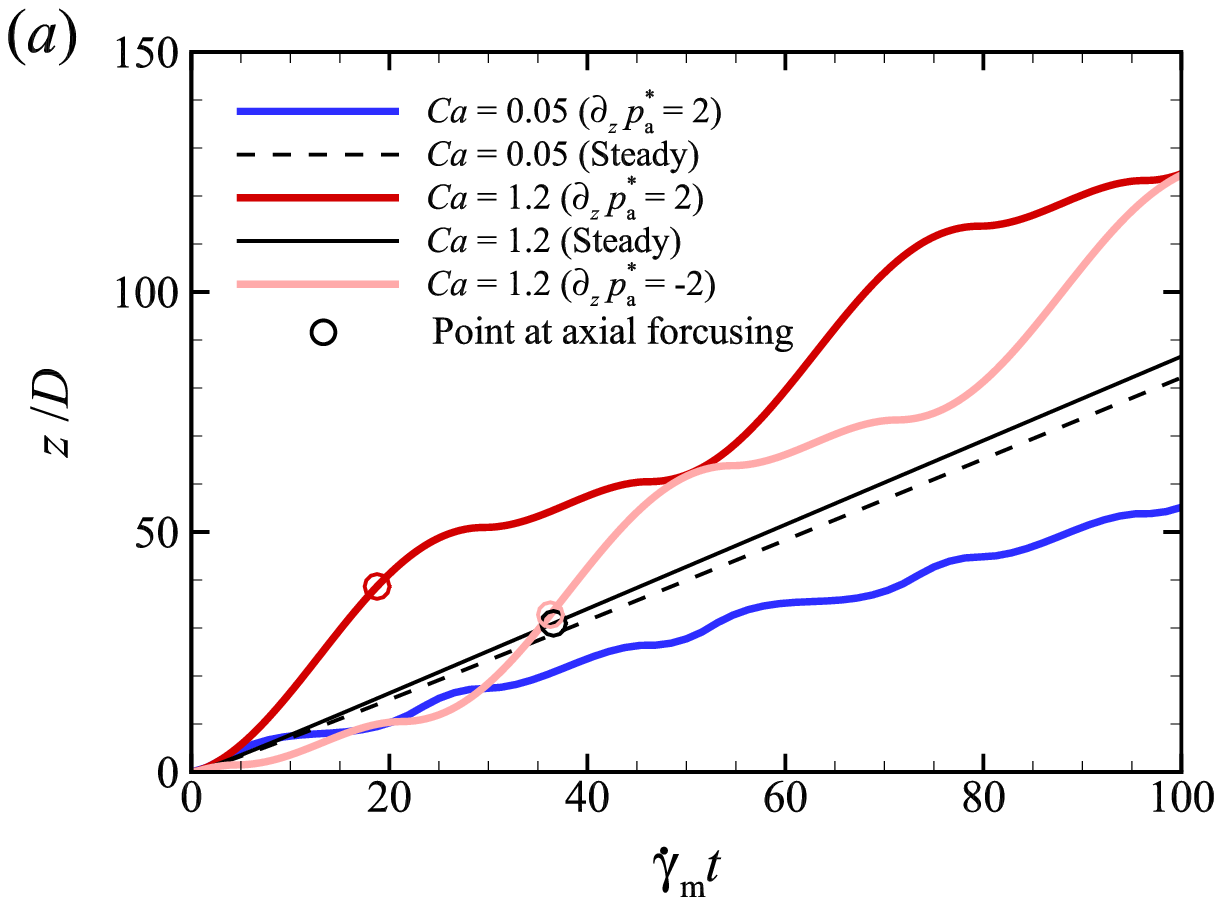}
  \includegraphics[height=4.5cm]{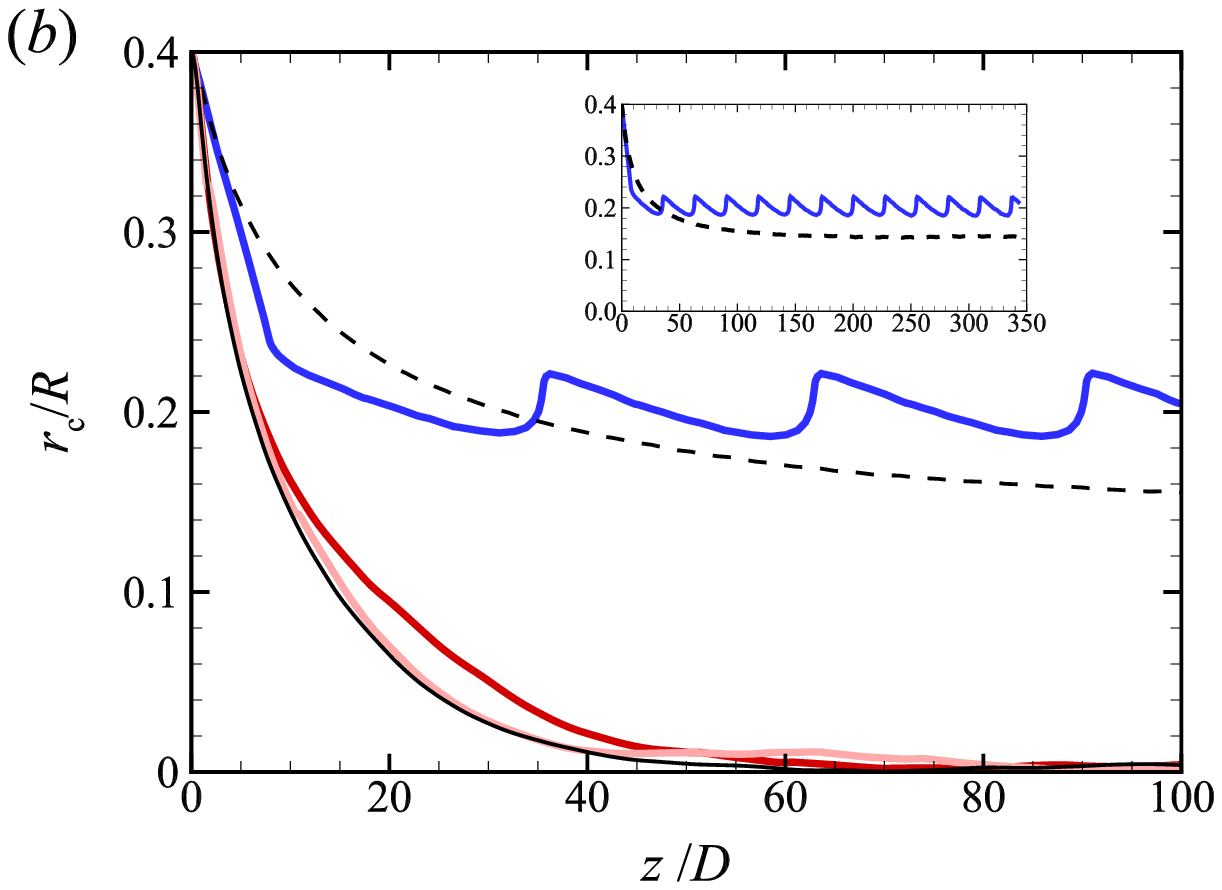}
  \caption{
         ($a$) Time history of the distance traveled along the flow direction ($z$-axis) $z/D$ in the case shown in figure~\ref{fig:re10f002},
         where the circle dots represent the points when the capsule has completed the axial focusing.
         ($b$) The radial position of capsule centroids $r_\mathrm{c}/R$ as a function of $z/D$.
 	  %The results are obtained with $Re = 10$, $R/a_0 = 2.5$, $r_0/R = 0.4$, and $\lambda = 1$.
  }
  \label{fig:re10f002_channel_length}
\end{figure}

Figure~\ref{fig:re10f002_channel_length}($a$) is the time history of the distance travelled along the flow direction ($z$-axis) $r_z/D$.
The distance to complete the axial focusing ($Ca = 1.2$) under pulsatile flow increases comparing to that in steady flow because the capsule speed along the flow direction increases by adding flow pulsation,
where the circle dots represent the points when the capsule has completed the axial focusing.
The capsule speed along the flow direction at $Ca = 0.05$, on the other hand, decreases with the pulsation of the flow.
Figure~\ref{fig:re10f002_channel_length}($b$) shows again the radial position of capsule centroids $r_\mathrm{c}/R$ as a function of $z/D$.
The capsule trajectories obtained for $Ca = 1.2$ remains almost the same,
while the capsule trajectory for $Ca = 0.05$ reaches equilibrium within a shorter traveled distance with pulsation.
%XXXFROM HEREXXX Although previous study by \cite{Vishwanathan2021} showed that inertial focusing of rigid sphere is achieved in only a fraction of the channel length ($1$ to $10$\%) compared to what would be required in a steady flow,
%it could be found especially for large pulsation amplitude, i.e., $\partial_z p_a/\partial_z p_0 \gg 1$,
%resulting in pulsation dominant flow of $V^\infty_\mathrm{osci}/V^\infty_0 \gg 1$,
%where $V^\infty_0 (= V_\mathrm{max}^\infty)$ and $V^\infty_\mathrm{osci}$ are the centreline (or maximum) velocity in the steady parabolic flow and oscillatory flow in the absence of any capsule. XXXTO HEREXXX
Following the classification by~\cite{Vishwanathan2021},
our problem is oscillatory dominated,
since the oscillation amplitude is one order of magnitude greater than the steady flow component (i.e., $O(s\omega/\bar{u}^\prime) \sim 10^1$,
where $s$ is the centreline displacement amplitude and $\bar{u}^\prime$ is the centreline velocity in a steady flow component).
Notwithstanding this, the oscillatory motion was not enough to enhance the inertial focusing,
in terms of channel lengths needed for the inertial focusing,
because of the capsule deformations impeding the inertial focusing,
%Our numerical results further showed that the channel length cannot be substantially shortened even when the pulsation is greater than steady component with $V^\infty_\mathrm{osci}/V^\infty_0 = 4\pi$, because the capsule deformations could impede off-centre focusing,
consistently with previous numerical study (see figure~4$a$ in~\cite{Takeishi2022}).

\begin{figure}%figure 7
  \centering
  \includegraphics[height=5.5cm]{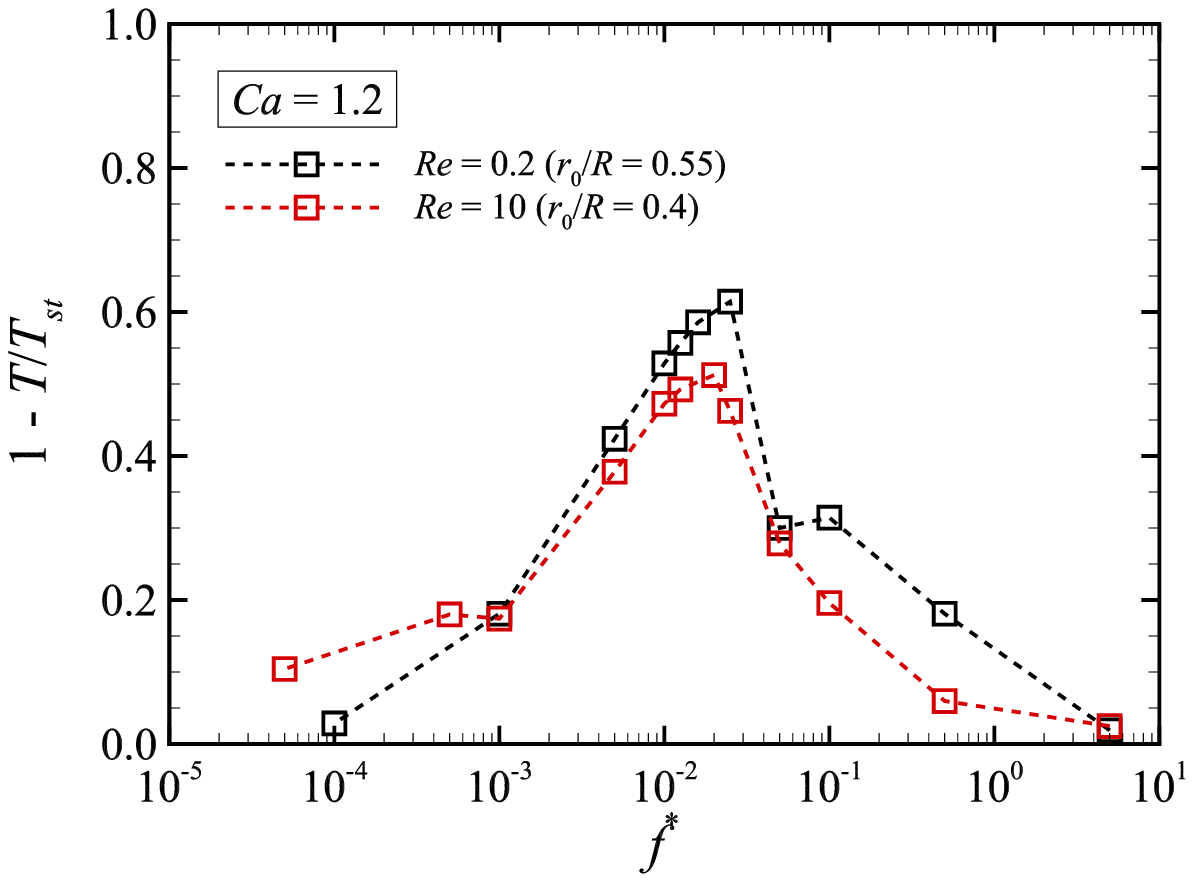}
  \caption{
         Acceleration indicator of the axial focusing $[1 - T/T_\mathrm{st}]$ as a function of the oscillatory frequency $f^\ast$ for different $Re$ (= 0.2 and 10).
         $T$ and $T_\mathrm{st}$ are the elapsed time needed by the capsule centroid to reach the centreline under pulsatile and steady flows,
         respectively.
         The initial radial position of the capsule is set to be $r^\ast_\mathrm{c0} = 0.55$ for $Re = 0.2$ (see also figure~4$a$ in~\cite{Takeishi2023}) and $r^\ast_\mathrm{c0} = 0.45$ for $Re = 10$.
         The results are obtained with $Ca$ = 1.2.
  }
  \label{fig:time}
\end{figure}

We now focus on axial focusing (i.e., cases of relatively high $Ca$) at finite $Re$.
As discussed in figure~\ref{fig:re10f002}($d$),
previous study showed that the speed of the axial focusing can be accelerated by the flow pulsation~\citep{Takeishi2023}. 
%\erase{We define the focusing times $T$ and $T_\mathrm{st}$ needed by the capsule centroid to reach the centreline (within a one fluid mesh corresponding to $\sim 6\%$ of its radius to account for the oscillations in the capsule trajectory) under pulsatile and steady flows, respectively.}
An acceleration indicator of the axial focusing $[1 - T/T_\mathrm{st}]$ at $Re = 10$ is summarised in figure~\ref{fig:time},
as a function of $f^\ast$ ($= f/\dot\gamma_\mathrm{m}$),
where the results at $Re = 0.2$~\citep{Takeishi2023} are also supperposed.
Although the initial radial position of the capsule $r^\ast_\mathrm{c0}$ is slightly different between the two $Re$,
the focusing time is commonly minimised at a specific frequency in both cases.
Note that, the values of the dimensional frequency depend on the estimation of $G_s$,
which varies with the membrane constitutive laws and which is also sensitive to different experimental methodologies, e.g., atomic force microscopy, micropipette aspiration, etc.~\citep{Bao2003};
the estimation of the dimensional frequency is therefore not trivial. %and left as for future investigations.
We hereby conclude that capsules with large $Ca$ exhibit axial focusing even at finite $Re$, and that their equilibrium radial positions are not altered by the flow pulsation.

We speculate that the optimal focusing frequency of $f^\ast \approx 0.02$,
corresponding to dimensional frequency of $f = 20$ Hz,
is the membrane resonance frequency,
given a reference radius of $a_0 = 4$ $\mu$m and the surface shear elastic modulus of $G_s = 4$ $\mu$N/m~\citep{Takeishi2014}.
However, there is currently no clear theoretical framework on the resonance frequency of capsule.
To provide further insights into the state of resonance,
we constructed a 2D fluid membrane model (or hydrodynamic equations of bilayer membrane),
obtained by Onsager's variation principle,
wherein the fluid membrane is assumed to be an almost planar bilayer membrane~\citep{Takeishi2024_ResultsEng}.
%Simulations were performed for a wide range of oscillatory frequency and membrane tensions.
Our numerical results showed that membrane characteristic shift from an elastic-dominant to viscous-dominant state appears within the range $40$ Hz $\leq f \leq 400$ Hz, almost independently of surface tensions (Figure 5c in~\cite{Takeishi2024_ResultsEng}).
Since the resonance frequency can be formulated with intrinsic material (membrane) properties,
it is expected that the value remains the same even under multi-capsule interactions.
Indeed, we discovered that crossover frequency of the storage and loss moduli in suspension of biconcave capsules modeling red blood cells (RBCs),
whose inverse is defined as a relaxation time,
is almost $40$ Hz, regardless of the volume fraction of the capsules (Fig. 7f, in~\cite{Takeishi2024}).
However note that, the critical frequency was commonly estimated in terms of order of magnitude ($O(f) = 10$ Hz) both in single and multi-capsule dynamics as well as theoretical principles,
sine its exact estimation depends on $G_s$.
Our recent numerical-experimental estimation strategy allows to quantify $G_s$ of intact RBCs under dynamics and derive its value as $\sim 0.5$ $\mu$N/m~\citep{Takeishi2024_JRheol},
which is one order of magnitude smaller than that obtained by the stretch test~\citep{Takeishi2014}.
Consequently, the dimensional optimal focusing frequency becomes $O(f) = 10^2$ Hz,
which is still in the range of the critical frequency estimated by the 2D fluid membrane model~\citep{Takeishi2024_ResultsEng}.
These results form a fundamental basis for further studies on resonance frequency of plasma membrane.

\subsection{Effect of Reynolds number on capsule behaviour under pulsatile channel flow}
\begin{figure}%figure 8
  \centering
  \includegraphics[height=4.5cm]{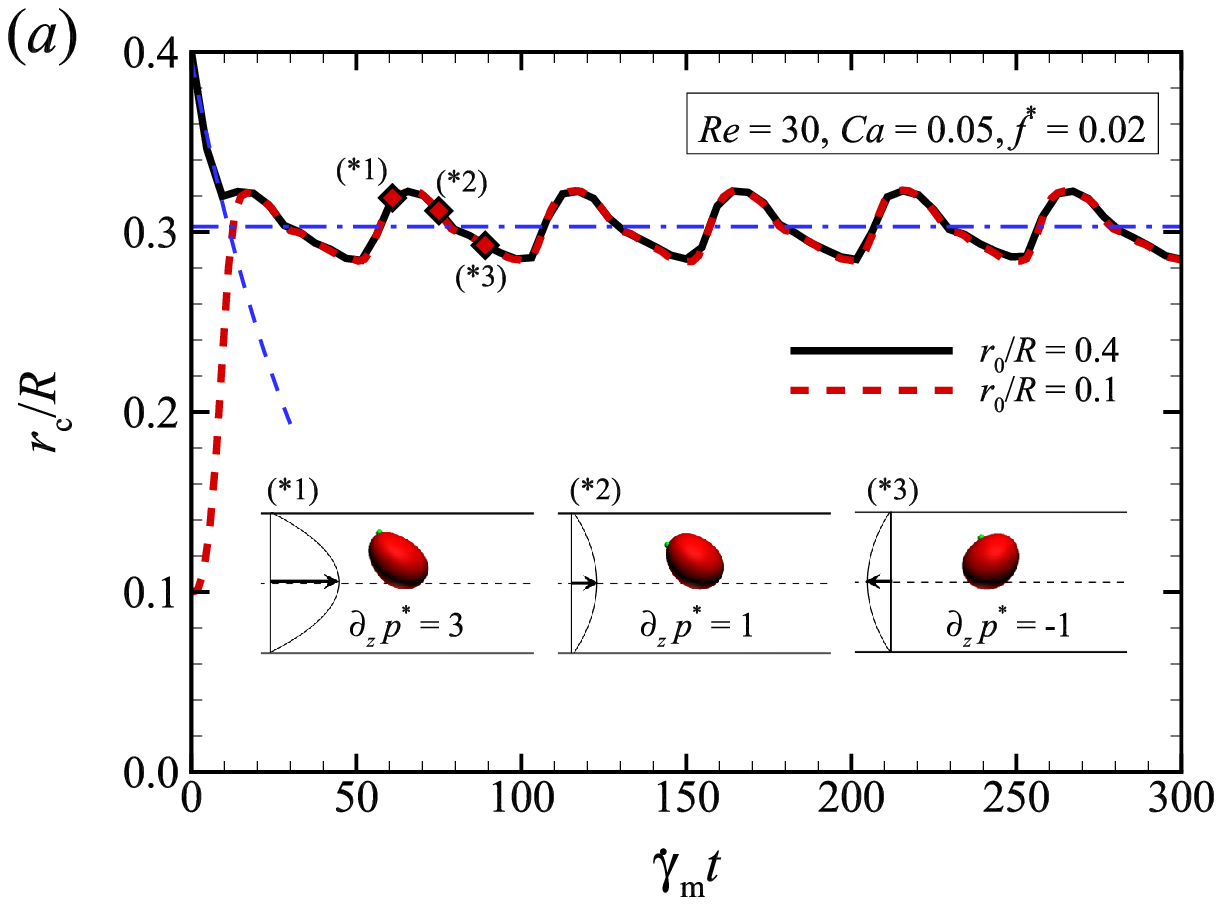}
  \includegraphics[height=4.5cm]{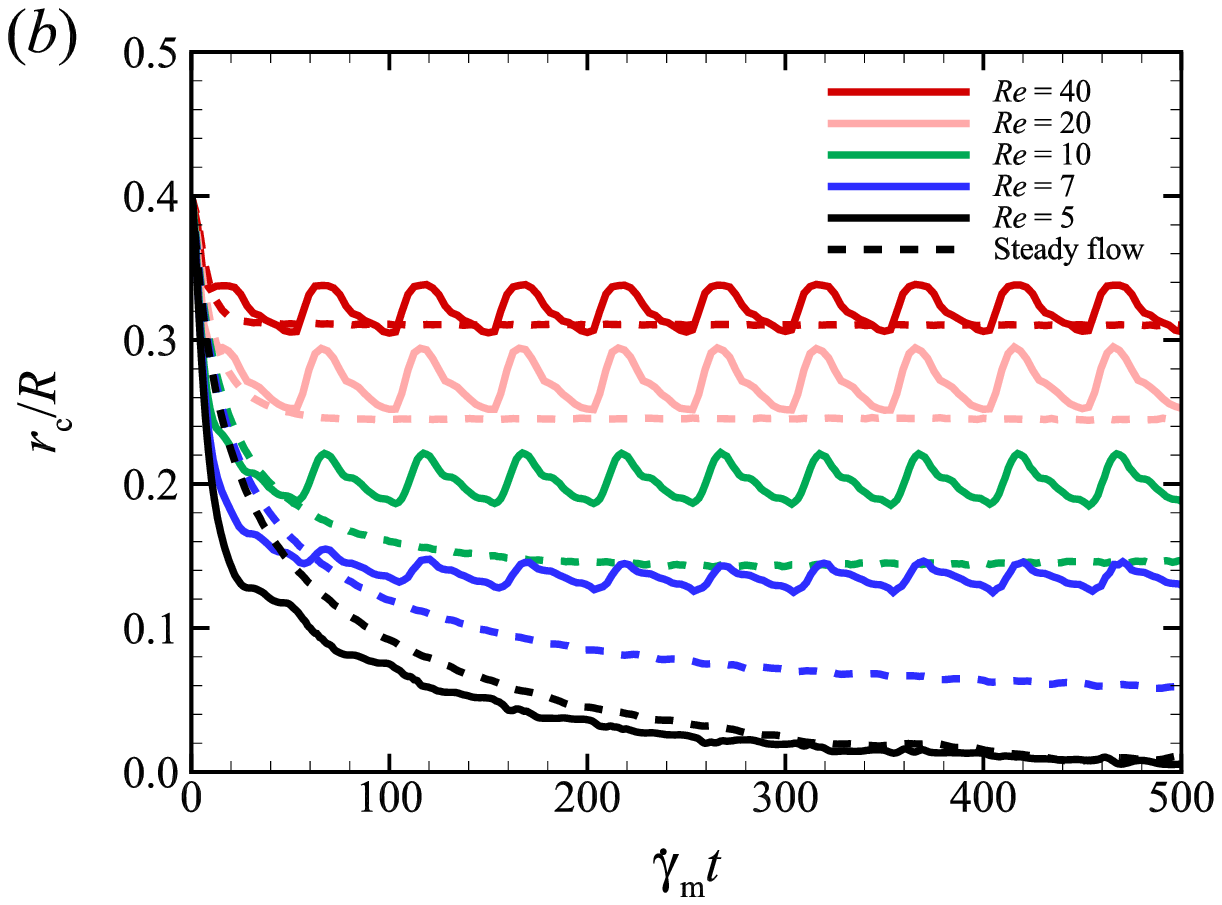}
  \includegraphics[height=4.5cm]{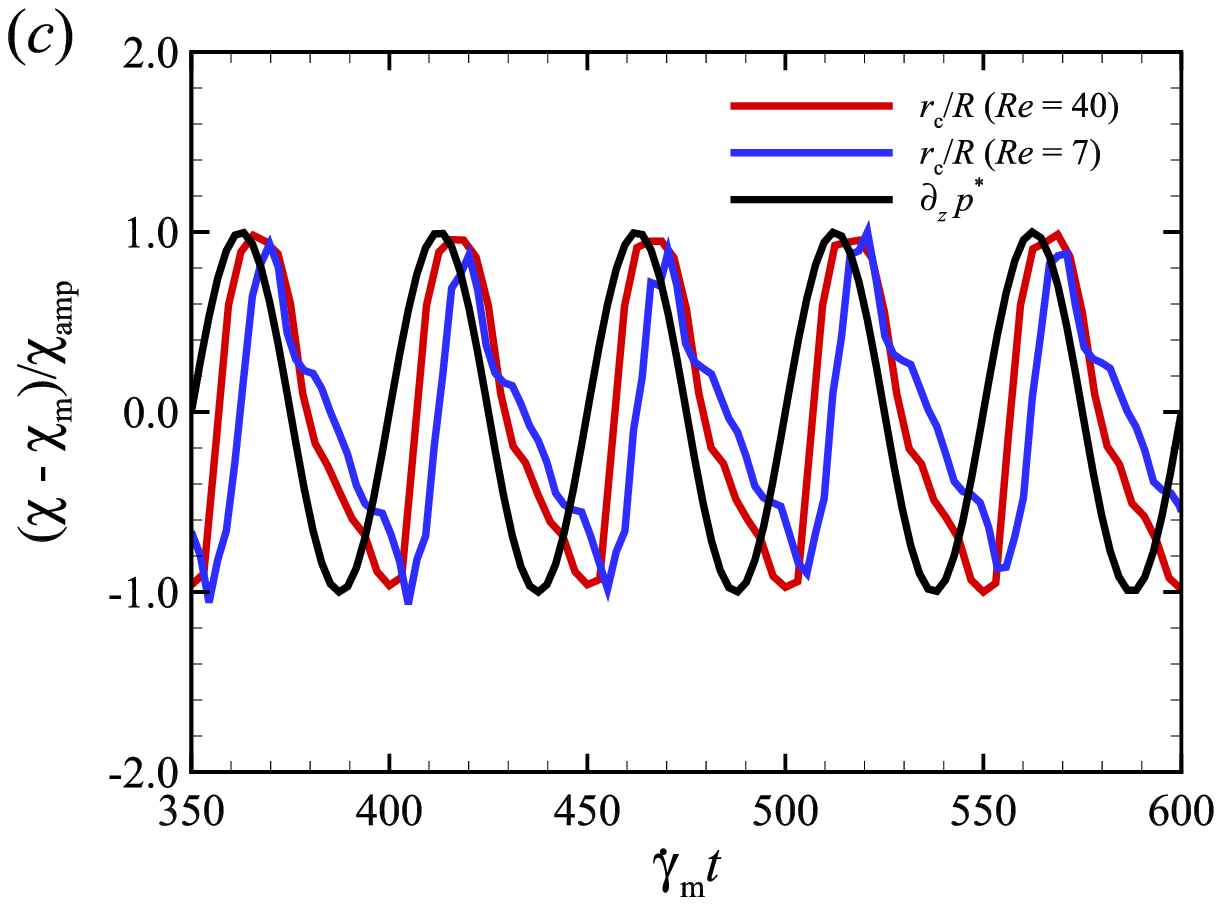}
  \includegraphics[height=4.5cm]{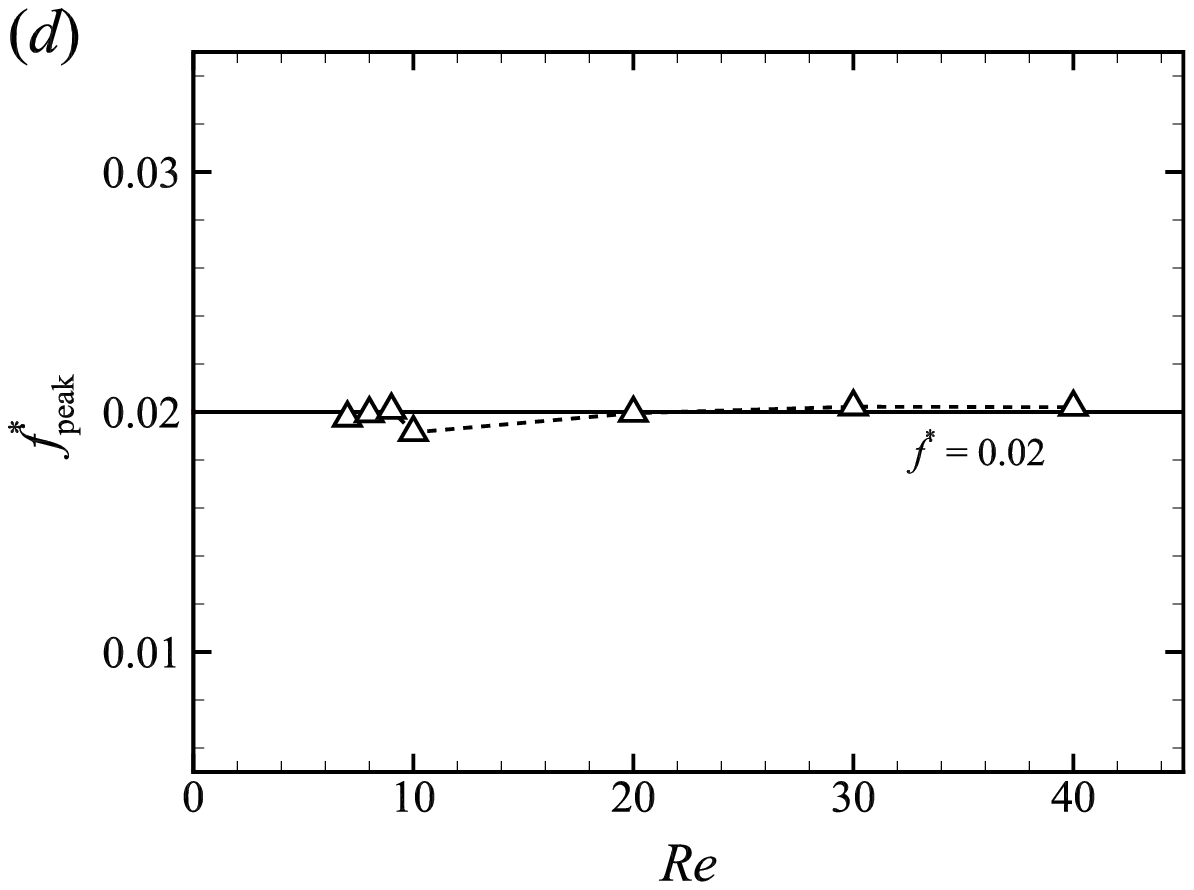}
  \caption{
         ($a$) Time histories of the radial position of the capsule centroid $r_\mathrm{c}/R$ at $Re = 30$ and $f^\ast = 0.02$ for different initial positions $r^\ast_\mathrm{c0}$ ($= 0.1$ and $0.4$),
         where insets represent snapshots of the lateral view of the deformed capsule at $\dot{\gamma}_\mathrm{m} t = 60$ ($^\ast$1), $75$ ($^\ast$2), and $90$ ($^\ast$3), respectively.
         Dashed lines are the curves $r_\mathrm{c}^\ast = C_2 \exp{(-C_1 t^\ast)}$,
         and the dash-dot line denotes the equilibrium radial position of the capsule centroid.
         ($b$) Time histories of $r_\mathrm{c}/R$ for different $Re$,
         where dashed lines denote those in steady flow.
         ($c$) Time histories of $r_\mathrm{c}/R$ and $\partial_z p^\ast$ at $Re = 7$ (blue) and $Re = 40$ (red),
         where the values are normalised by the amplitude $\chi_\mathrm{amp}$, and are shifted so that each baseline is the corresponding mean value $\chi_\mathrm{m}$.
         Data is shown after $\dot\gamma_\mathrm{m} t > 300$.
         ($d$) The peak frequency $f^\ast_\mathrm{peak}$ of the capsule centroid $r_\mathrm{c}/R$.
         %and ($d$) the phase delay $\delta^\ast$ from the pressure gradient $\partial_z p^\ast$, as a function of $Re$.
         The solid line in panel ($c$) denotes the oscillatory frequency $f^\ast = 0.02$.
 	  The results are obtained with $Ca = 0.05$, $R/a_0 = 2.5$, and $r^\ast_\mathrm{c0} = 0.4$.
  }
  \label{fig:correlation}
\end{figure}

We now focus on the inertial focusing of  capsules at relatively small $Ca$, and, unless otherwise specified, we show the results obtained for $Ca = 0.05$.
Figure~\ref{fig:correlation}($a$) shows representative time history of the capsule centroid during inertial (or off-centre) focusing at $Re = 30$ and $f^\ast = 0.02$ for different initial position of the capsule $r^\ast_\mathrm{c0}$ ($= 0.1$ and $0.4$),
where insets represent snapshots of the lateral view of deformed capsule at various time $\dot{\gamma}_\mathrm{m} t$ ($= 60$, $75$, and $90$), respectively.
The results clearly show that the equilibrium radial position of the capsule is independent of its initial position $r_0$
(except when $r_0 = 0$ for which the capsule remains at centreline).
Hereafter, each run case is started from a slightly off-centre radial position $r^\ast_\mathrm{c0} = 0.4$ ($R/a_0 = 2.5$).
For the trajectory at early times ($\dot{\gamma}_\mathrm{m} t \leq$ 20),
fitting by equation~\eqref{fitting} still works.
At quasi-steady state ($\dot{\gamma}_\mathrm{m} t >$ 20),
the capsule centroid fluctuates around an off-centre position $r_\mathrm{c}/R$ ($\approx 0.3$).
Thus, the trajectory of the capsule during inertial focusing can be expressed as
\begin{equation}
  r_\mathrm{c}^\ast = 
  \begin{cases}
  C_2 \exp{(-C_1 t^\ast)} & \text{for} \ t^\ast \leq t_\mathrm{ax}^\ast \\
  r_\mathrm{e}^\ast + \Delta r_\mathrm{osci}^\ast & \text{for} \ t^\ast > t_\mathrm{ax}^\ast
  \end{cases}
  ,
  \label{estimation}
\end{equation}
where $t_\mathrm{ax}^\ast$ is the time period during axial focusing,
$r^\ast_\mathrm{e}$ is the equilibrium radial position of the capsule centroid due to inertia,
and $\Delta  r_\mathrm{osci}^\ast$ is a perturbation due to the oscillatory flow.
Here, the equilibrium radial position is measured numerically by time averaging the radial position of the capsule centroid as $r^\ast_\mathrm{e} = \langle r_\mathrm{c}^\ast \rangle = \left( 1/\m{T} \right) \int_{t^\ast}^{t^\ast + \m{T}} r_\mathrm{c} (t^\prime) dt^\prime$.
%, where $\m{T}$ is the period of oscillation of the trajectory.

Figure~\ref{fig:correlation}($b$) shows the time histories of the capsule centroid $r_\mathrm{c}/R$ at $f^\ast = 0.02$ for different $Re$,
together with those with steady flow.
We observe that the radial positions are greater than those at steady flow for all $Re$, due to the larger values achieved by the pressure gradient during the pulsation.
However, the actual contribution of the oscillatory flow to the inertial focusing depends on $Re$.
For instance, for $Re \leq 7$, the capsule exhibits axial focusing at steady flow,
but a pulsatile channel flow allows the capsule to exhibit off-centre focusing.
Therefore, the pulsation itself can impede the axial focusing.

Figure~\ref{fig:correlation}($c$) shows the waveforms of $r_\mathrm{c}/R$ at the end of the migration ($\dot\gamma_\mathrm{m} \geq 350$),
where the instantaneous values are  normalised by their respective amplitudes $\chi_\mathrm{amp}$ and and shifted so that each baseline is the mean value $\chi_\mathrm{m}$.
Although the delay of $r_\mathrm{c}/R$ from the oscillatory pressure gradient $\partial_z p^\ast$ tends to decrease as $Re$ increases,
the overall waveforms of $r_\mathrm{c}/R$ well follow that of $\partial_z p^\ast$, as shown in figure~\ref{fig:re10f002}($b$).
To quantify the waveform of $r_\mathrm{c}/R$ and its correlation to $\partial_z p^\ast$,
we extract the dominant (or peak) frequency $f^\ast_\mathrm{peak}$ of $r_\mathrm{c}/R$ with a discrete Fourier transform,
whose principle and implementation are described in~\cite{Takeishi2024},
and the result are shown as a function of $Re$ in figures~\ref{fig:correlation}($d$).
In the cases of $Re \leq 6$, the capsule does not exhibit off-centre focusing, and thus the plots are displayed for $Re \geq 7$ only.
The value of $f^\ast_\mathrm{peak}$ collapses on the frequency of $\partial_z p^\ast$ with $f^\ast = 0.02$ for $Re \geq 7$ (figure~\ref{fig:correlation}$d$).
The transition from the axial focusing to the off-centre focusing thus requires a synchronisation, induced by capsule deformability, between the capsule centroid and the background pressure gradient.

\begin{figure}%figure 9
  \centering
  \includegraphics[height=4.5cm]{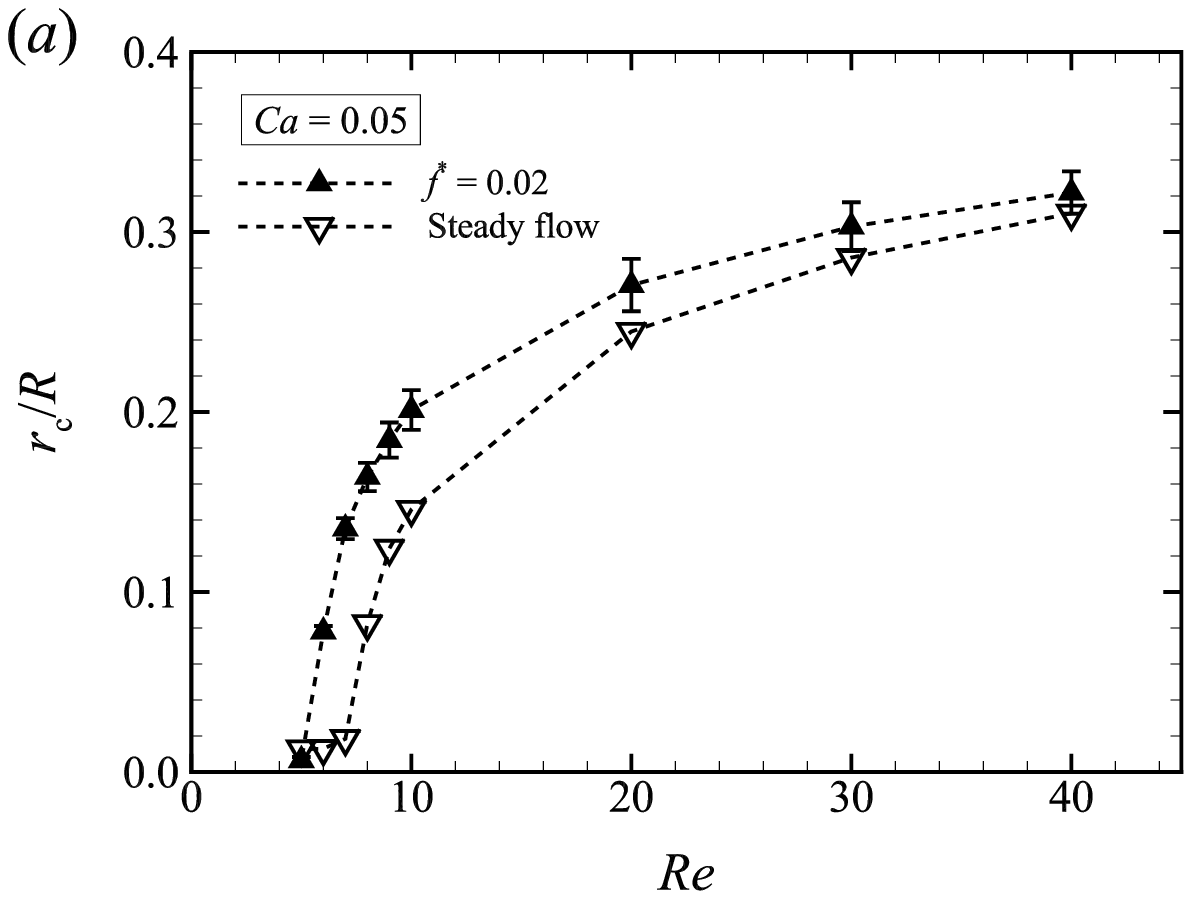}
  \includegraphics[height=4.5cm]{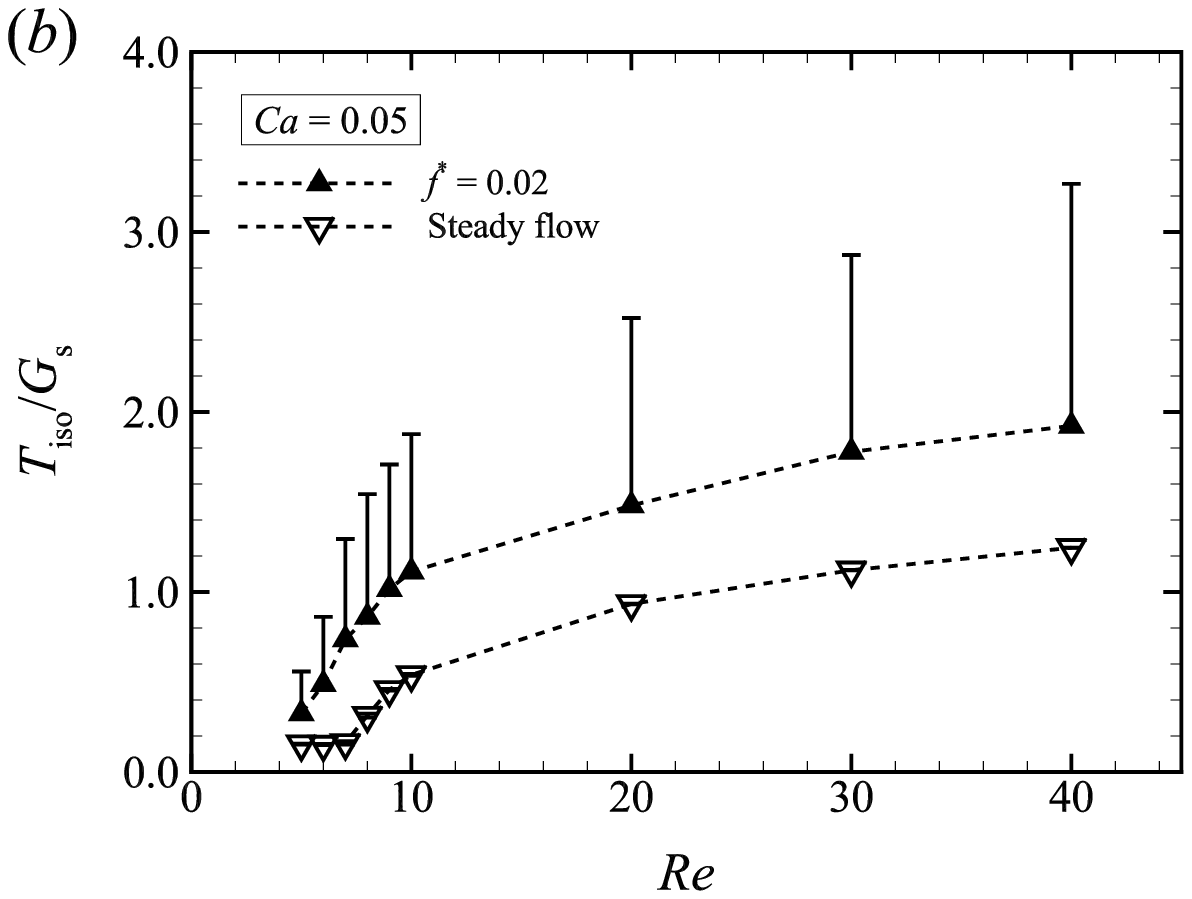}
  \caption{
         Time average of ($a$) the radial position of the capsule centroid $\langle r_\mathrm{c} \rangle/R$,
         and ($b$) isotropic tensions $\langle T_\mathrm{iso} \rangle$ as a function of $Re$ at $f^\ast$ = 0.02,
         where the error bars represent the standard deviation during a period.
         The error bars in panel ($b$) are displayed only on one side of the mean value for major clarity. 
 	  The results are obtained with $Ca$ = 0.05, $R/a_0$ = 2.5, and $r^\ast_\mathrm{c0} = 0.4$.
  }
  \label{fig:effect_re}
\end{figure}

Figures~\ref{fig:effect_re}($a$) and \ref{fig:effect_re}($b$) show the time average of the radial position or equilibrium position $\langle r_\mathrm{c} \rangle/R$ and the isotropic tension $\langle T_\mathrm{iso} \rangle$, respectively, as a function of $Re$,
where the error bars represent the standard deviation (SD) during a period.
%Here, the error bars in $\langle T_\mathrm{iso} \rangle$ are displayed only on one side of the mean value for major clarity.
Overall, both these values nonlinearly increase with $Re$,
with the mean values in the oscillating flows  always greater than those in steady flows.
The curves show steep increases for $Re \leq 10$, 
followed by a more moderate increases for $Re > 10$;
these general tendency are the same in steady or pulsatile flows.
The effect of the flow pulsation is maximised at moderate $Re$ ($= 7$),
in which the axial focusing is impeded by the pulsatile flow (figure~\ref{fig:effect_re}$a$).
%The results also suggest that significant contribution of pulsatile flow at relatively large $Re$ will require large oscillatory amplitude.
The results also show that small fluctuations of the capsule radial position ($SD(r_\mathrm{c}/R) < 10^{-2}$) are accompanied by large fluctuations of the membrane tension ($SD(T_\mathrm{iso}) > 10^{-1}$).

\subsection{Effect of oscillatory frequency on capsule behaviour under pulsatile channel flow}
\begin{figure}%figure 10
  \centering
  \includegraphics[height=4.5cm]{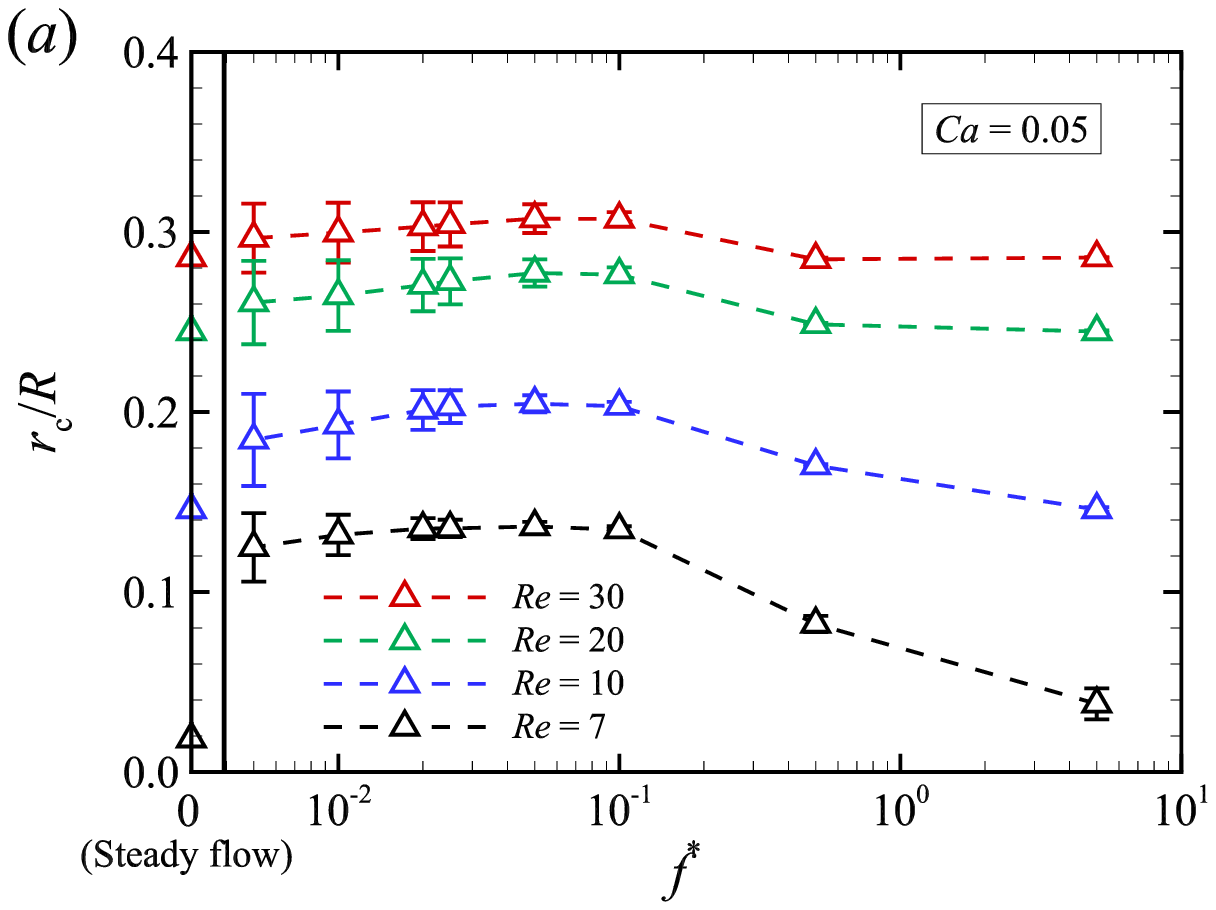}
  \includegraphics[height=4.5cm]{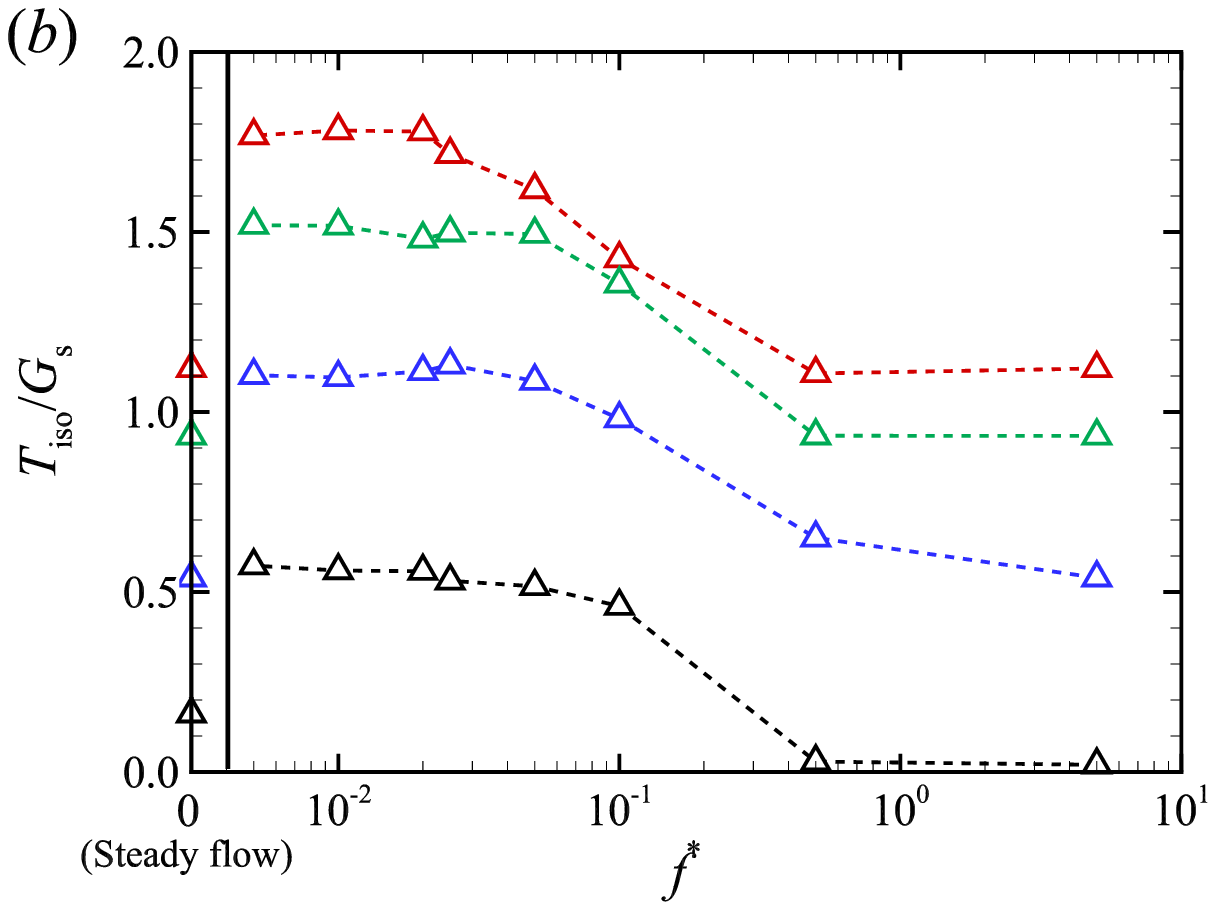}
  \caption{
         Time average of ($a$) $\langle r_\mathrm{c} \rangle/R$ and ($b$) $\langle T_\mathrm{iso} \rangle/R$ as a function of $f^\ast$.
         The error bars in panel ($b$) are not displayed for major clarity. 
         All results are obtained with $R/a_0 = 2.5$, $Re = 10$, and $Ca = 0.05$.
  }
  \label{fig:effect_f}
\end{figure}

Finally, we investigate the effect of the oscillatory frequency $f^\ast$ on the equilibrium radial position $\langle r_\mathrm{c} \rangle/R$ at $Re = 10$,
with the results summarised in Figure~\ref{fig:effect_f}($a$),
where those at steady flow are also displayed at the point $f^\ast = 0$.
The results clearly suggest that there exists a specific frequency to maximise $\langle r_\mathrm{c} \rangle/R$, independently of $Re$.
Interestingly, such effective frequency ($f^\ast$ = 0.05) are close to or slightly larger than those maximising the axial focusing speed (see figure~\ref{fig:time}).
Comparing to steady flow,
the equilibrium radial position $\langle r_\mathrm{c} \rangle/R$ at the effective frequency was enhanced by 640\% at $Re$ = 7, 40\% at $Re$ = 10, 13\% at $Re$ = 20, and 7.6\% at $Re$ = 30.
The contribution of the oscillatory flow to the off-centre focusing becomes negligible for higher frequencies,
in which the trajectory of the capsule centroid at the highest frequency considered ($f^\ast = 5$) collapses on that obtained with steady flow.

%\mr{XXX QUI $\leftarrow$ What?}
Figure~\ref{fig:effect_f}($b$) shows the time average of the isotropic tension $\langle T_\mathrm{iso} \rangle$ as a function of $f^\ast$.
The values of $\langle T_\mathrm{iso} \rangle$ decrease as $f^\ast$ increases because of the reduction of the shear stress when moving closer to the channel centreline (i.e., small $\langle r_\mathrm{c} \rangle/R$).
The results of large capsule deformation at relatively small frequencies are consistent with a previous numerical study by~\cite{Matsunaga2015}, 
who showed that at high frequency a neo-Hookean spherical capsule undergoing oscillating sinusoidal shear flow cannot adapt to the flow changes,
and only slightly deforms, consistently with  predictions obtained by asymptotic theory~\citep{BarthesBiesel1981, BarthesBiesel1985}. Thus, capsules at low frequencies exhibit an overshoot phenomenon,
in which the peak deformation is larger than that its value in steady shear flow. %and increases with viscosity contrast $\lambda$ and the mean value of $Ca$.

\begin{figure}%figure 11
  \centering
  \includegraphics[height=4.5cm]{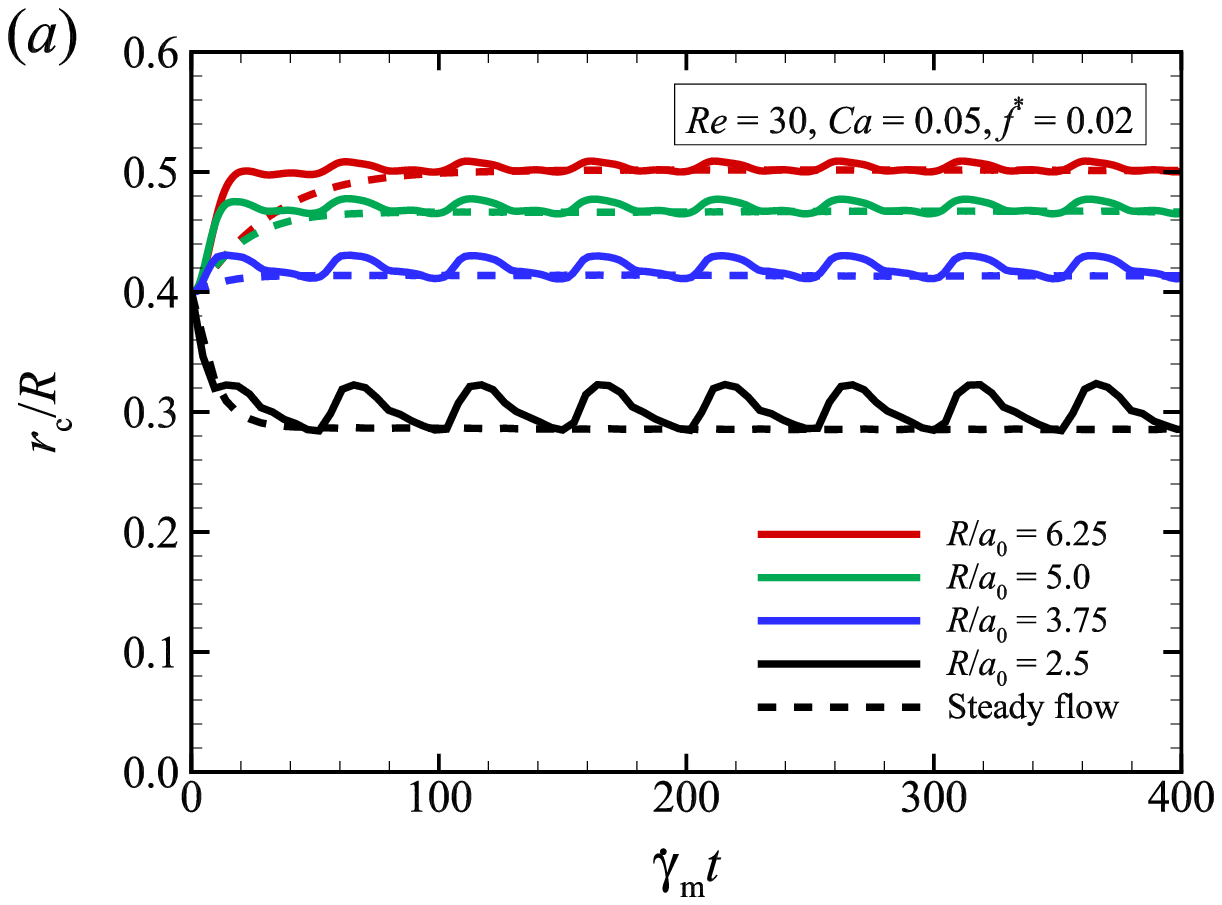}
  \includegraphics[height=4.5cm]{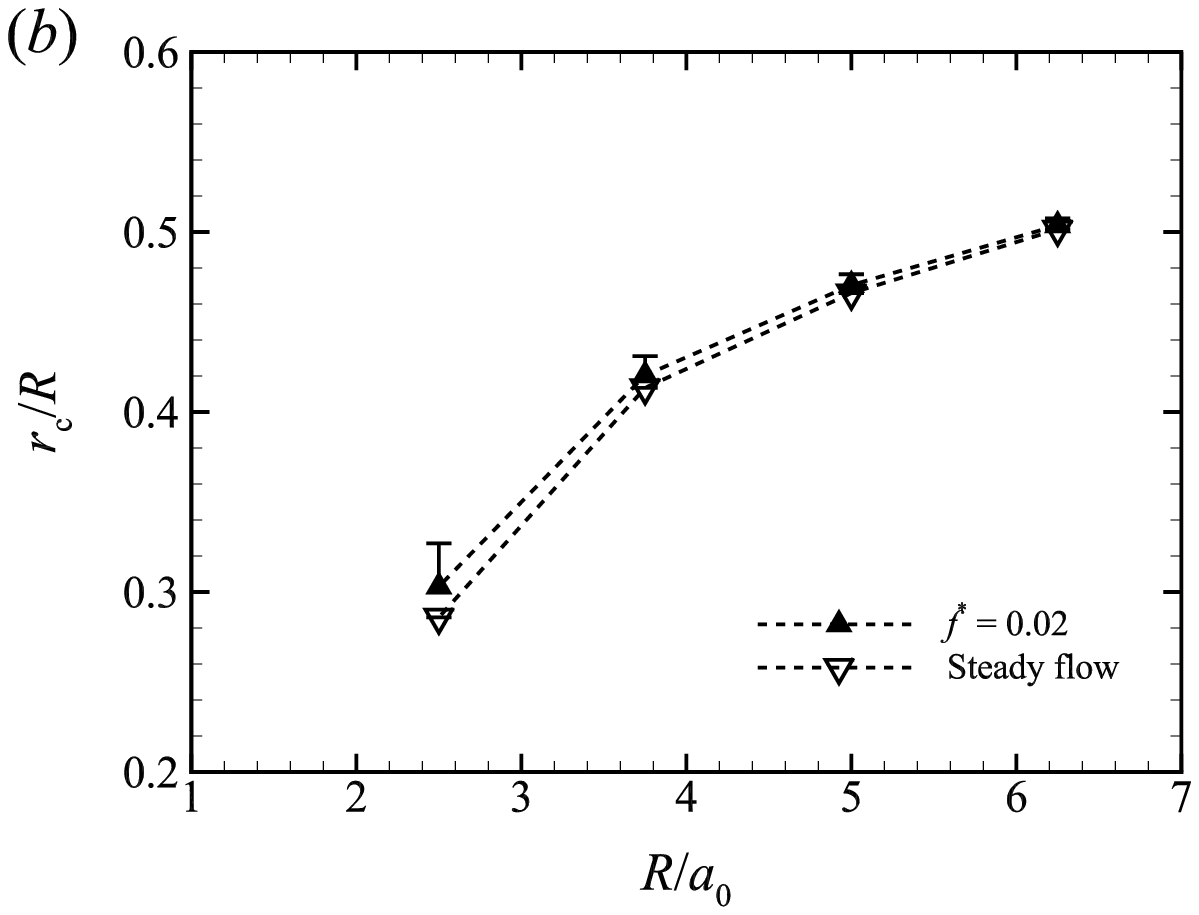}
  \includegraphics[height=4.5cm]{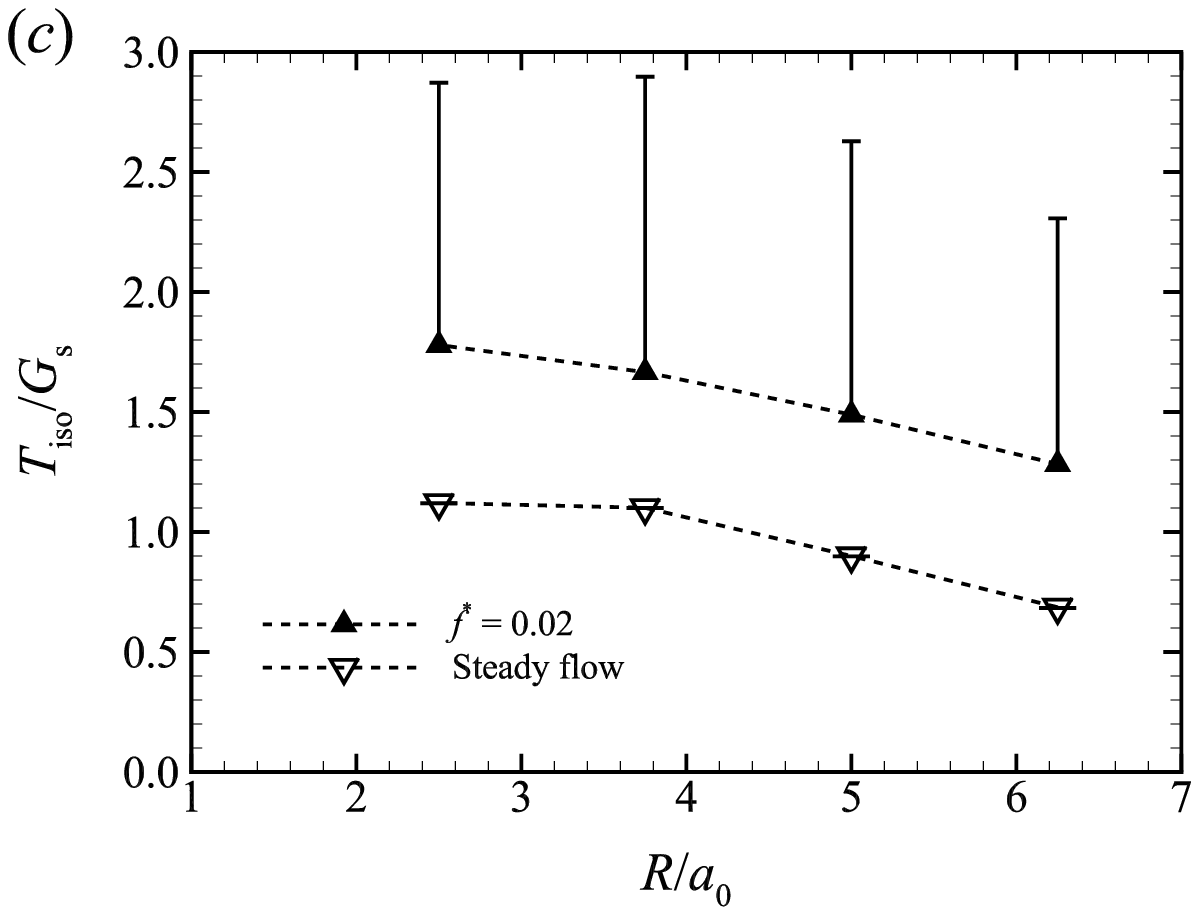}
  \includegraphics[height=4.5cm]{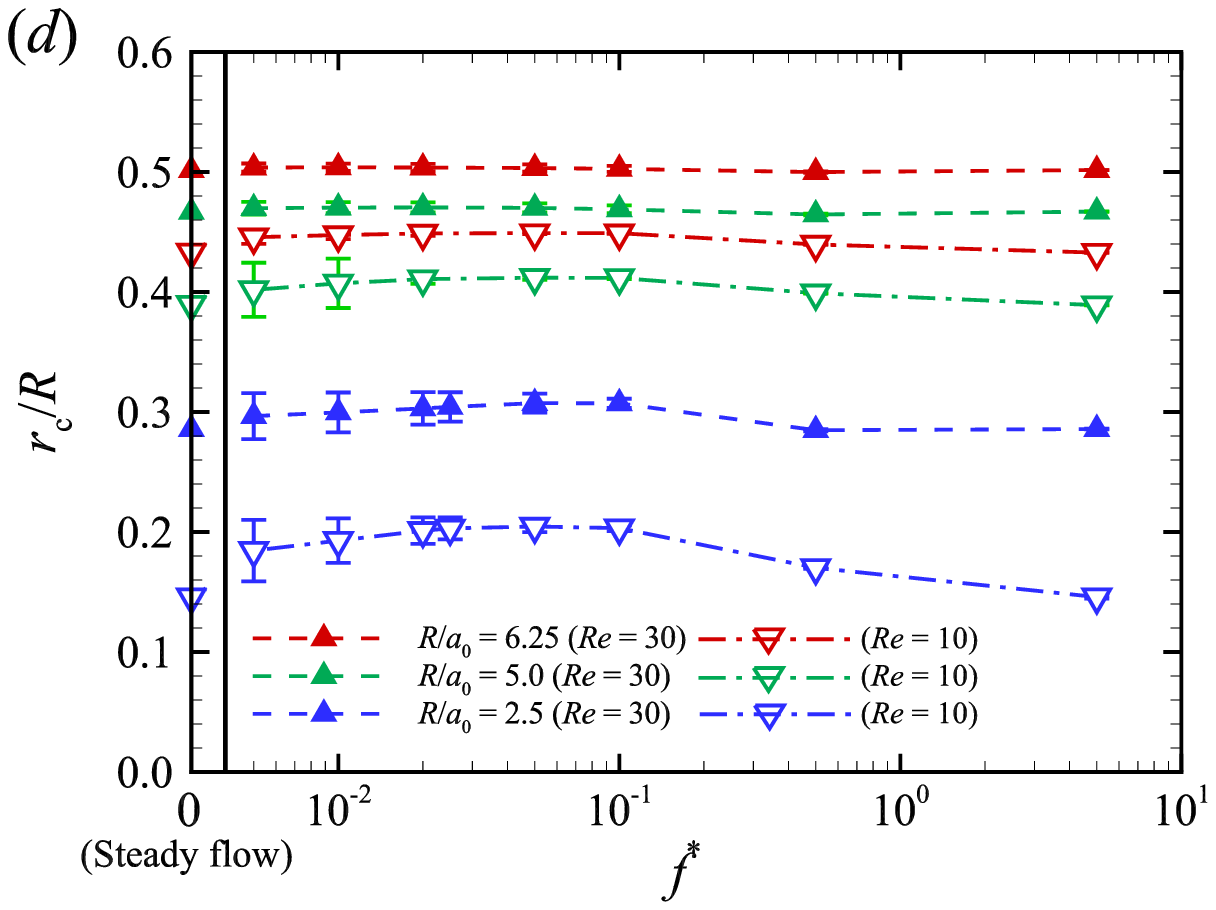}
  \caption{
         ($a$) Time history of $r_\mathrm{c}/R$ for different size ratios channel sizes $R/a_0$.
         ($b$ and $c$) Time average of ($b$) $\langle r_\mathrm{c} \rangle/R$, and ($c$) $\langle T_\mathrm{iso} \rangle/R$ as a function of $R/a_0$.
         The error bars in panels ($b$) and ($c$) are displayed only on one side of the mean value for major clarity. 
         ($d$) Time average of $\langle r_\mathrm{c} \rangle/R$ at $Ca$ = 0.05 as function of $f^\ast$ for different $R/a_0$.
         All results are obtained with $Re$ = 30, $Ca$ = 0.05, and $f^\ast$ = 0.02,
         and data at $Re = 10$ is superposed on the panel ($d$).
  }
  \label{fig:effect_D}
\end{figure}

By increasing channel diameter $D$ (= $2R$ = 30 $\mu$m, 40 $\mu$m, and 50 $\mu$m),
we also investigate the effect of the size ratio $R/a_0$ ($= 3.75$, $5$, and $6.25$) on the equilibrium radial position $\langle r_\mathrm{c} \rangle/R$.
Figure~\ref{fig:effect_D}($a$) is the time history of $r_\mathrm{c}/R$ for different size ratios $R/a_0$ at $Re = 30$, and $f^\ast = 0.02$,
where the trajectories obtained with the steady flow are also displayed.
All run cases are started from $r^\ast_\mathrm{c0} = 0.4$.
The equilibrium radial positions increase with $R/a_0$,
while the contribution of oscillatory flow to $\langle r_\mathrm{c} \rangle/R$ becomes small as well as its fluctuation.
This is quantified in figure~\ref{fig:effect_D}($b$),
where $\langle r_\mathrm{c} \rangle/R$ is shown as a function of the size ratio $R/a_0$.
%\nt{An increase in the equilibrium radial position $\langle r_\mathrm{c} \rangle/R$ with $R/a_0$ is consistent with previous 2D pulsatile channel flow~\cite{Sun2009}.}
Although the equilibrium radial position $\langle r_\mathrm{c} \rangle/R$ increases with $R/a_0$,
indicating that dimensional equilibrium radial position $\langle r_\mathrm{c} \rangle$ also increases with $R$,
the isotropic tension $\langle T_\mathrm{iso} \rangle/G_s$ decreases as shown in figure~\ref{fig:effect_D}($c$).
This is because the distance from the capsule centroid to the wall ($R - \langle r_\mathrm{c} \rangle$) increases with $R$, resulting in lower shear stress.
Oscillatory-dependent off-centre focusing is summarised in figure~\ref{fig:effect_D}($d$),
where the results are obtained with different channel size $R/a_0$ and different $Re$ ($= 10$ and $30$).
The result shows that oscillatory-dependent off-centre focusing is impeded as $Re$ increases.

It is known that rigid particles align in an annulus at a radius of about $0.6R$ for $Re = D\overline{V}/\nu = O(1)$~\citep{Segre1962, Matas2004, Matas2009},
and shift to larger radius for larger $Re$~\citep{Matas2004, Matas2009},
where $\overline{V}$ is the average axial velocity~\citep{Matas2004}.
Our numerical results show that capsules with low deformability ($Ca = 0.05$) are still in $\langle r_\mathrm{c} \rangle/R \sim 0.5$ even for the largest channels ($R/a_0$ = 6.25; $R$ = 25 $\mu$m) and Reynolds numbers ($Re = 30$), both in the steady and pulsatile flows (figure~\ref{fig:effect_D}$b$).
Therefore, off-centre focusing is impeded even at such small particle deformability.
This result is consistent with previous numerical study about a spherical hyperelastic particle in a circular channel with $R/a_0$ = 5 under steady flow for 100 $\leq Re \leq$ 400 and 0.00125 $\leq We \leq$ 4~\citep{Alghalibi2019}.
There, the authors showed that the particle radial position is $\langle r_\mathrm{c} \rangle/R \sim 0.5$ at the highest $Re$ ($= 400$) and lowest $We$ ($= 0.00125$).
Our numerical results further show that the contribution of the flow pulsation to the off-centre focusing decreases as the channel size $R/a_0$ increases (figures~\ref{fig:effect_D}$b$ and \ref{fig:effect_D}$d$) because of the low shear stress acting on the membranes (figure~\ref{fig:effect_D}$c$).
In other word, a large amplitude is required for oscillaton-induced off-centre focusing in high $Re$ and large channels.

%In previous numerical study by~\cite{Sun2009},
%the behaviour of neutrally buoyant circular particle in oscillatory pressure-driven channel flows were conceptually explained by background 2D channel flow profile.
The Poiseuille flow in a rigid circular pipe subject to the action of an oscillation pressure gradient was well described in~\citet{Uchida1956, Womersley1955}.
At low frequency, oscillatory flow in the tube is better able to keep pace with the changing pressure.
In the limit of zero frequency,
the relation between flow and pressure is instantaneous as in a steady Poiseuille flow (see figure~\ref{fig:vprofile}$a$ in Appendix~\S\ref{appA_vprof}).
Thus, the particle (or capsule) is subject to shear stress,
which results in its lateral movement due to the shear-induced lift force at low frequency.
We confirmed it in Figure~\ref{fig:effect_f}($a$),
which is also consistent with the 2D numerical results of a neutrally buoyant circular particle \citep{Sun2009}.
However, since the mechanism of axial focusing of capsules is primarily attributed to their deformability, the frequency-dependent axial focusing of rigid (circular) particle remains unclear.
At high frequency, on the other hand, oscillatory flow in a channel is less able to keep pace with the changing pressure,
thus reaching less than the fully developed Poiseuille flow profile (almost flat velocity profile) at the peak of each cycle (see figure~\ref{fig:vprofile}$b$ in Appendix~\S\ref{appA_vprof}).
In the limit of infinite frequency, the velocity reached at the peak of each cycle is zero,
that is, the fluid does not move at all.
Thus, the particle does not experience shear stress and maintains its lateral position at high frequency, consistently with the 2D numerical analysis by~\cite{Sun2009}.

Furthermore, we showed that $\langle r_\mathrm{c} \rangle/R$ increased with $Re (\leq 30)$, results consistent with those of rigid spherical particles in three-dimensional (3D) steady pipe flows~\citep{Sun2009}.
Such an increase in $\langle r_\mathrm{c} \rangle/R$ with $Re$ can also be found for a rigid spherical particle on the square channel face,
especially for $Re \leq 100$~\citep{Nakagawa2015},
and also observed experimentally by~\cite{Miura2014, Choi2011, Abbas2014}.
It is also known that the channel face equilibrium positions decrease with $Re$ in particular for $Re > 100$, while the channel corner equilibrium positions continue to increase with $Re$~\citep{Nakagawa2015}.
Although our numerical results described in Figure~\ref{fig:effect_D}($d$) suggest that a large amplitude is required for oscillaton-induced off-centre focusing at high $Re$,
it remains an open question whether the off-center focusing of capsules can indeed be enhanced by large-amplitude pulsatile flow and whether the optimal frequency remains consistent with the value measured in this study ($f^\ast \approx 0.05$).
%Therefore, a discrepancy in the equilibrium position described in~\citep{Sun2009} compared to our numerical results was attributed to differences in channel geometries (2D channel or 3D tube flow) and $Re$.

Throughout our analyses, we have quantified the radial position of the capsule in a tube based on the empirical expression~\eqref{estimation}.
%\begin{equation}
%  r_\mathrm{c}^\ast = 
%  \begin{cases}
%  C_2 \exp{(-C_1 t^\ast)} & \text{for} \ t^\ast \leq t_\mathrm{ax}^\ast \\
%  r_\mathrm{e}^\ast + \Delta r_\mathrm{osci}^\ast & \text{for} \ t^\ast > t_\mathrm{ax}^\ast
%  \end{cases}
%  ,
%  \label{estimation}
%\end{equation}
%where $t_\mathrm{ax}^\ast$ is the time period during axial focusing.
We have provided insights about the coefficient $C_1$ ($> 0$) in $r_\mathrm{c}^\ast = C_2 \exp{(-C_1 t^\ast)}$, which potentially scales the lift force and depends on shape, i.e., capillary number $Ca$ and viscosity ratio $\lambda$.
%The formulating lift force as functions of these parameters is not trivial and is our future study.

\section{Conclusion}
We numerically investigated the lateral movement of spherical capsules in steady and pulsatile channel flows of a Newtonian fluid, for a wide range of $Re$ and oscillatory frequency $f^\ast$.
The roles of size ratio $R/a_0$, and capillary number $Ca$ on the lateral movement of the capsule have been evaluated and discussed.
The first important question we focused on is whether a capsule lateral movement at finite $Re$ in a pulsatile channel flow can be altered by its deformability.
The second question is whether equilibrium radial positions or traveling time are controllable by oscillatory frequency.

Our numerical results showed that capsules with high $Ca$ still exhibit axial focusing even at finite $Re$ (e.g., $Re = 10$), and that their equilibrium radial positions cannot be altered by flow pulsation.
However, the speed of axial focusing at such high $Ca$ is substantially accelerated by making the driving pressure gradient oscillating in time.
We also confirmed that there exists a most effective frequency ($f^\ast \approx 0.02$) which maximises the speed of axial focusing, and that it remains the same as that in almost inertialess condition.
For relatively low $Ca$, on the other hand, the capsule exhibits off-centre focusing, resulting in an equilibrium radial position $\langle r_\mathrm{c} \rangle/R$ which depends on $Re$.
There also exists a specific frequency to maximise $\langle r_\mathrm{c} \rangle/R$, which is independent of $Re$.
Interestingly, such effective frequency ($f^\ast \approx  0.05$) is close to that for axial focusing.

Frequency-dependent inertial focusing requires a synchronisation between the radial centroid position of the capsule and the background pressure gradient, resulting in periodic and large membrane tension, which impedes axial focusing.
Such synchronisation abruptly appear at $O(Re) = 10^0$,
and shifts to an almost perfect syncrohisation as $Re$ increases.
Thus, there is almost no contribution of flow pulsation to $\langle r_\mathrm{c} \rangle/R$ at relatively low $Re$ ($\leq 5$) or large $Re$ ($\geq 30$),
while the contribution of the pulsation to $\langle r_\mathrm{c} \rangle/R$ is maximised at moderate $Re$ ($\approx 7$),
allowing the capsule to exhibit axial focusing in steady flow.
For constant amplitude of oscillatory pressure gradient,
oscillatory-dependent inertial focusing is impeded as $Re$ and channel diameter increase,
and thus relatively large oscillatory amplitude is required in such high $Re$ and large channels.
Throughout our analyses, we have quantified the radial position of the capsule in a tube based on the empirical expression. We hereby conclude that the knowledge obtained under inertialess conditions~\citep{Takeishi2023} has been extended to cases involving finite $Re$ ($> 1$) conditions.

Given that the speed of inertial focusing can be controlled by oscillatory frequency,
the results obtained here can be utilised for label-free cell alignment/sorting/separation techniques, e.g., for circulating tumor cells in cancer patients or precious hematopoietic cells such as colony-forming cells.

\section*{Acknowledgements}
This research was supported %by JSPS KAKENHI Grant Numbers \nt{JP20H04504}, and
by the Okinawa Institute of Science and Technology Graduate University (OIST) with subsidy funding to M.E.R. from the Cabinet Office, Government of Japan.
The presented study was partially funded by Daicel Corporation. 
K.I. acknowledges the Japan Society for the Promotion of Science (JSPS) KAKENHI for Transformative Research Areas A (Grant No. 21H05309) and the Japan Science and Technology Agency (JST), FOREST (Grant No. JPMJFR212N).
M.E.R. also acknowledges funding from the Japan Society for the Promotion of Science (JSPS), grants 24K17210 and 24K00810.
%\nt{Finally, the collaborative research was supported by the SHINKA grant provided by OIST.}

\section*{Conflicts of Interest}
The authors report no conflict of interest.

%\section*{Supplementary movie}
%Supplementary movies are available at \textit{https://doi.org/xxx.yyy.zzz}.

\appendix
\section{\label{appA1}Numerical setup and verification}
\begin{figure}%figure 12
  \centering
  \includegraphics[height=5.5cm]{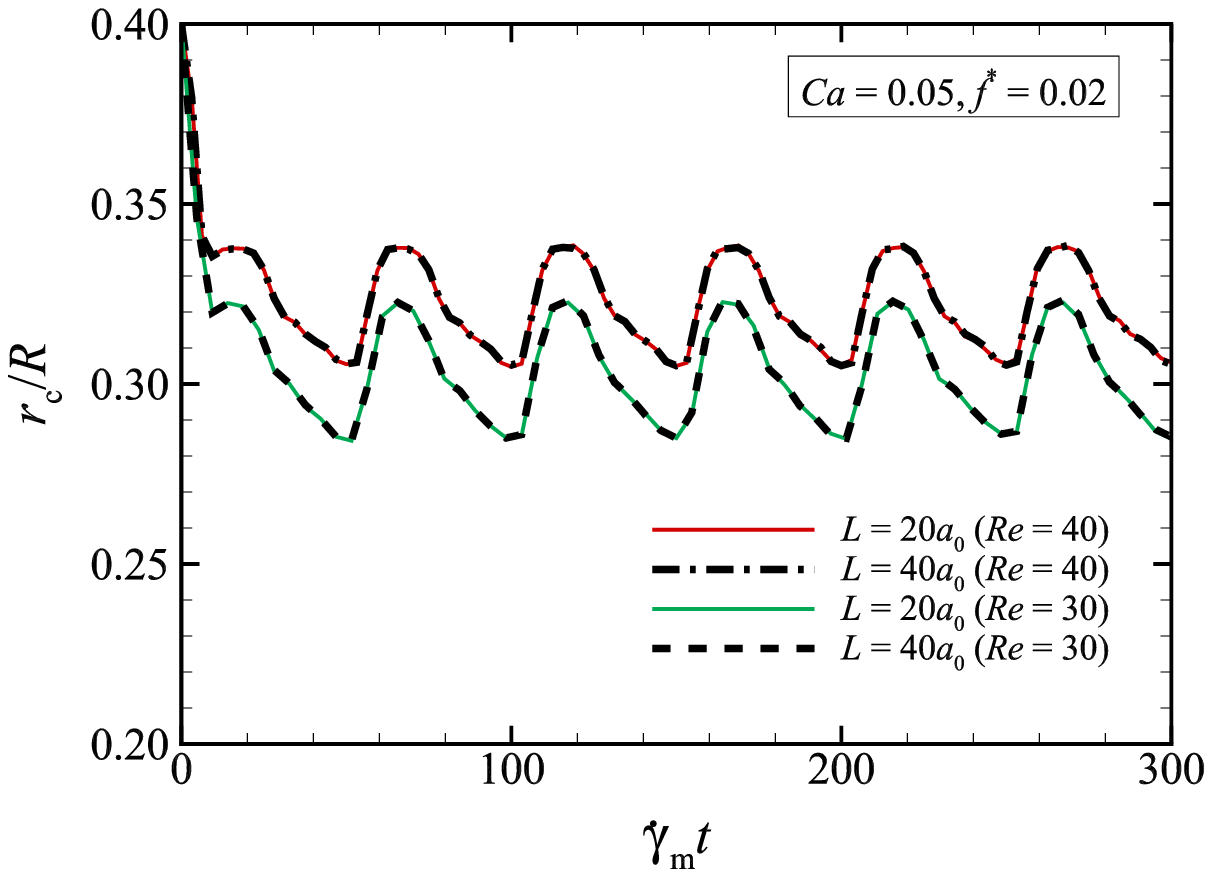}
  \caption{
         Time history of the radial position $r/R$ for different channel lengths $L$ ($= 20 a_0$ and $40 a_0$) and different $Re$ ($= 30$ and $40$).
         %($b$) Time history of $r/R$ at $Re$ = 30 for different different mesh resolutions (250 $\mu$m/lattice and 125 $\mu$m/lattice).
         %($b$) Time history of $r/R$ at $Re$ = 7 in $L$ = 20$a_0$.
         %\nt{The trajectory obtained with the double mesh resolution (125 $\mu$m/lattice) are also shown.}
         In all runs, the capsule is initially placed at $r^\ast_\mathrm{c0} = 0.4$.
         The results are obtained with $R/a_0 = 2.5$, and $Ca = 0.05$.
  }
  \label{fig:verification}
\end{figure}

To show that the channel length is adequate for studying the behaviour of a capsule that is subject to inertial flow,
we have tested the channel length $L$ ($= 20a_0$ and $40a_0$),
and investigated its effect on the radial positions of the capsule centroids.
The time history of the radial position of the capsule centroid $r_\mathrm{c}$ is compared between these different channel lengths in figure~\ref{fig:verification}, where the centroid position $r_\mathrm{c}$ is normalised by the channel radius $R$. 
The results obtained with the channel length $L$ used in the main work ($= 20 a_0$) are consistent with those obtained with twice longer channel ($L = 40 a_0$).
%We have also confirmed the equilibrium radial position of the capsule centroid for $Re$ = 7,
%where the channel length remains the same as $L$ = 20$a_0$.
%\nt{Figure~\ref{fig:verification}($b$) shows the time history of $r/R$ at $f^\ast$ = 0.02 and steady flow.
%The results are consistent with those obtained with double mesh resolution (125 $\mu$m/lattice).}

\section{\label{appA_Lift}Lift force on a capsule in a Poiseuille flow}
We consider an object immersed in a Poisseulle flow,
assuming that the flow is in the (steady) Stokes regime and that the object size is much smaller than the channel size.
We also neglect any boundary effects acting on the object.
Let $y$ be the position relative to the channel centre.
Due to the linearity of the Stokes equation, the object experiences a hydrodynamic resistance proportional to its moving velocity, given by
%In particular, we are interested in \ki {the time evolution of $y$. The hydrodynamic drag is written as}%lift force $f_1^L$ that centralise the object into the channel centre,
\begin{equation}
  f_1^L = - \xi_1 \dot{y}.
\end{equation}
%where the dot symbol indicates the time derivative.
Note that the drag coefficient $\xi_1 > 0$ is only determined by the viscosity and the shape (including the orientation) of the particle.
We then consider the effects of the background Poiseuille flow.
We have assumed that the channel size is much larger than the particle size, and hence the background flow to the particle is well approximated by a local shear flow with its local shear strength,
\begin{equation}
  \dot\gamma = -2 \frac{V_\mathrm{max}^\infty}{R^2} y.
\end{equation}
In the presence of the background shear,
the shear-induced lift force in general appears,
and this is proportional to the shear strength \citep{Kim-Karrila2005},
\begin{equation}
  f_2^L = -\xi_2 \dot\gamma = 2 \xi_2 \frac{V_\mathrm{max}^\infty}{R^2} y,
\end{equation}
where the coefficient $\xi_2$ is again only determined by the viscosity and the shape.
The force balance equation on the $y$-direction therefore reads $f_1^L + f_2^L = 0$.
If we introduce a new shape-dependent coefficient, $C_1$, as
\begin{equation}
  C_1 = 2 \frac{\xi_2}{\xi_1} \frac{V_\mathrm{max}^\infty}{R^2},
\end{equation}
we obtain the evolution equation for the position $y$ as
\begin{equation}
  \dot{y} = - C_1 y.
\end{equation}
This equation is easily solved if $C_1$ is constant and the result is the exponential accumulation to the channel centre, consistent with the numerical results. 

\section{\label{appA2}Neo-Hookean spherical capsule}
In consideration of previous works by, e.g.,~\cite{Lefebvre2007, Wang2021},
the trajectory of capsule centroids are compared between different types of membrane constitutive law for a comprehensive understanding of capsule motion in a tube, and to verify whether our empirical expression~\eqref{fitting} works independently of the membrane constitutive law.
Here, let us take the NK constitutive law, which is given by
\begin{equation}
  \frac{w_\mathrm{NH}}{G_s} = \frac{1}{2} \left( I_1 - 1 + \frac{1}{I_2 + 1} \right).
  \label{NH}
\end{equation}
\begin{figure}%figure 13
  \centering
  \includegraphics[height=6cm]{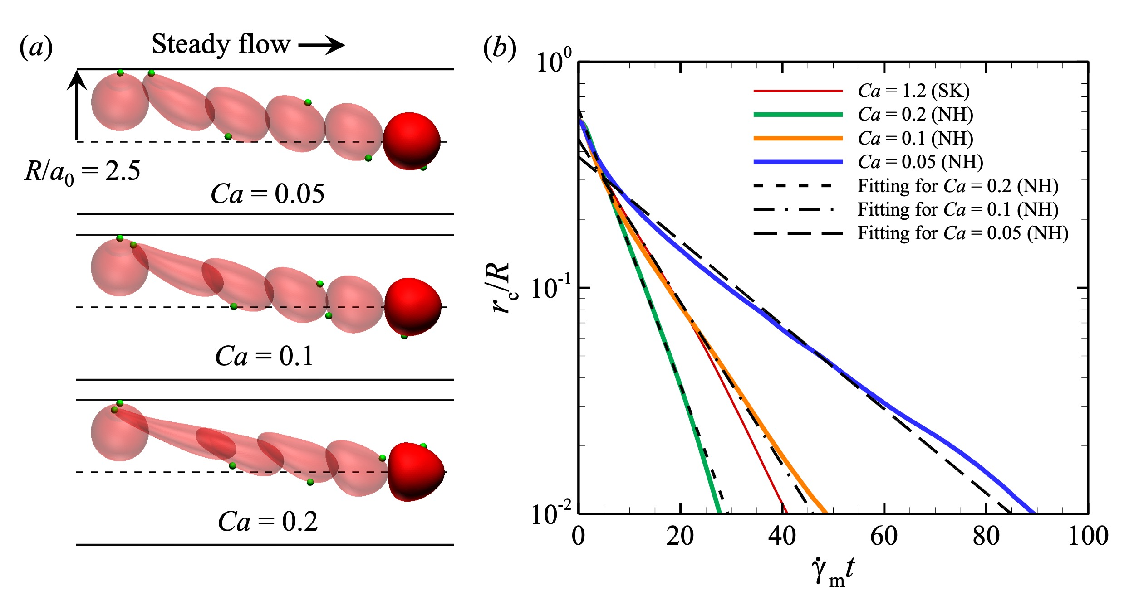}
  \caption{
         ($a$) Side views of the capsule during its axial focusing under steady flow for $Ca = 0.05$ (top), $Ca = 0.1$ (middle), and $Ca = 0.2$ (bottom).
         The capsule is initially placed at $r^\ast_\mathrm{c0} = 0.55$.
         ($b$) Time histories of the radial position of these capsule centroids $r_\mathrm{c}/R$.
         The dashed lines are the curves $r_\mathrm{c}^\ast = C_2 \exp{(-C_1 t^\ast)}$.
         The result at the highest $Ca$ ($= 1.2$) obtained with SK law is also superposed.
 	  The results are obtained with $Re= 0.2$, $R/a_0 = 2.5$, and $\lambda = 1$.
  }
  \label{fig:fitting_NH}
\end{figure}

Figure~\ref{fig:fitting_NH}($a$) shows side views of the capsule during its axial focusing at each time for different $Ca$ ($= 0.05$, $0.1$, and $0.2$).
Other numerical settings ($Re$, initial position $r^\ast_\mathrm{c0}$, and viscosity ratio $\lambda$) are the same as described in \S\ref{S1}.
Even at relatively small $Ca$ ($= 0.2$),
the NH-capsule exhibits large elongation after flow onsets,
resulting in fast axial focusing.
The trajectory and fitting for it at each $Ca$ are shown in figure~\ref{fig:fitting_NH}($b$),
where the result at the highest $Ca$ ($= 1.2$) obtained with SK law described in figure~\ref{fig:fitting}($b$) is also superposed.
The results suggest that equation~\eqref{fitting} still works even for NH-spherical capsules, although the applicable ranges of $Ca$ are relatively small compared to those described by the SK law.

\section{\label{appA3}Taylor parameter}
The SK-spherical capsule deformation is quantified by the Taylor parameter $D_{12}$, defined as
\begin{equation}
  D_{12} = \frac{| a_1 - a_2 |}{a_1 + a_2},
  \label{D12}
\end{equation}
where $a_1$ and $a_2$ are the lengths of the semi-major and semi-minor axes of the capsule, and are obtained from the eigenvalues of the inertia tensor of an equivalent ellipsoid approximating the deformed capsule~\citep{Ramanujan1998}.
\begin{figure}%figure 14
  \centering
  \includegraphics[height=5.5cm]{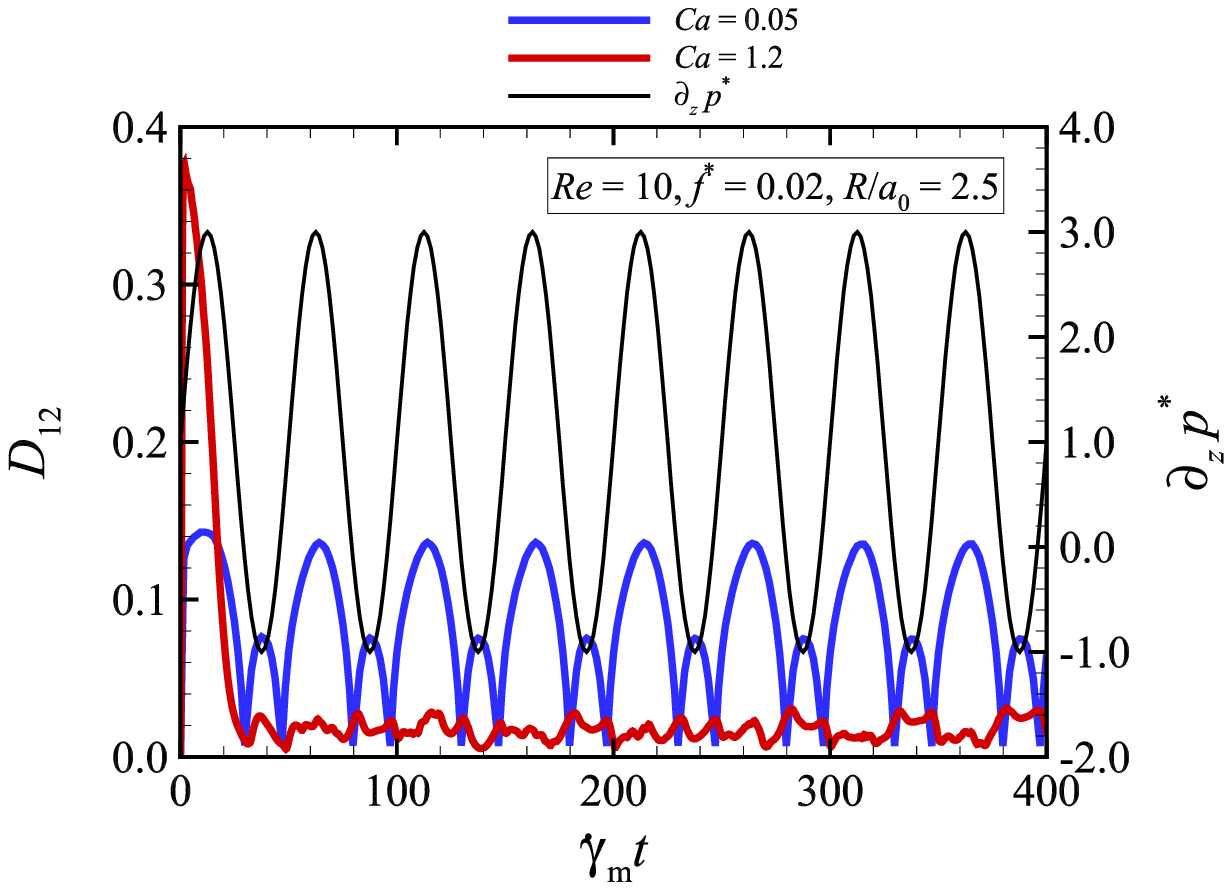}
    \caption{
         Time histories of the Taylor parameter $D_{12}$ for different $Ca$ ($= 0.05$ and $1.2$) at $Re = 0.2$.
 	  The results are obtained with $f^\ast = 0.02$, and $R/a_0 = 2.5$.
  }
  \label{fig:D12_time}
\end{figure}

Figure~\ref{fig:D12_time} shows the time history of $D_{12}$ at $Re = 10$, $R/a_0 = 2.5$, and $f^\ast = 0.02$.
Differently from what observed for the isotropic tension shown in figure~\ref{fig:re10f002}($c$),
the off-centred capsule exhibits large $D_{12}$,
which well responds to the oscillatory pressure $\partial_z p^\ast$. %,
%comparing to the axial-symmetric shaped capsule on the channel centre.
Thus, the magnitude of $D_{12}$ is strongly correlated with the capsule radial position (and the consequent shear gradient).
%Note that the lengths of semi-minor axes of deformed capsule $a_2$ are almost the same while the lengths of the semi-major axis $a_1$ are greater in $Ca = 0.05$ that in $Ca = 1.2$ (data not shown).

\begin{figure}%figure 15
  \centering
  \includegraphics[height=4.5cm]{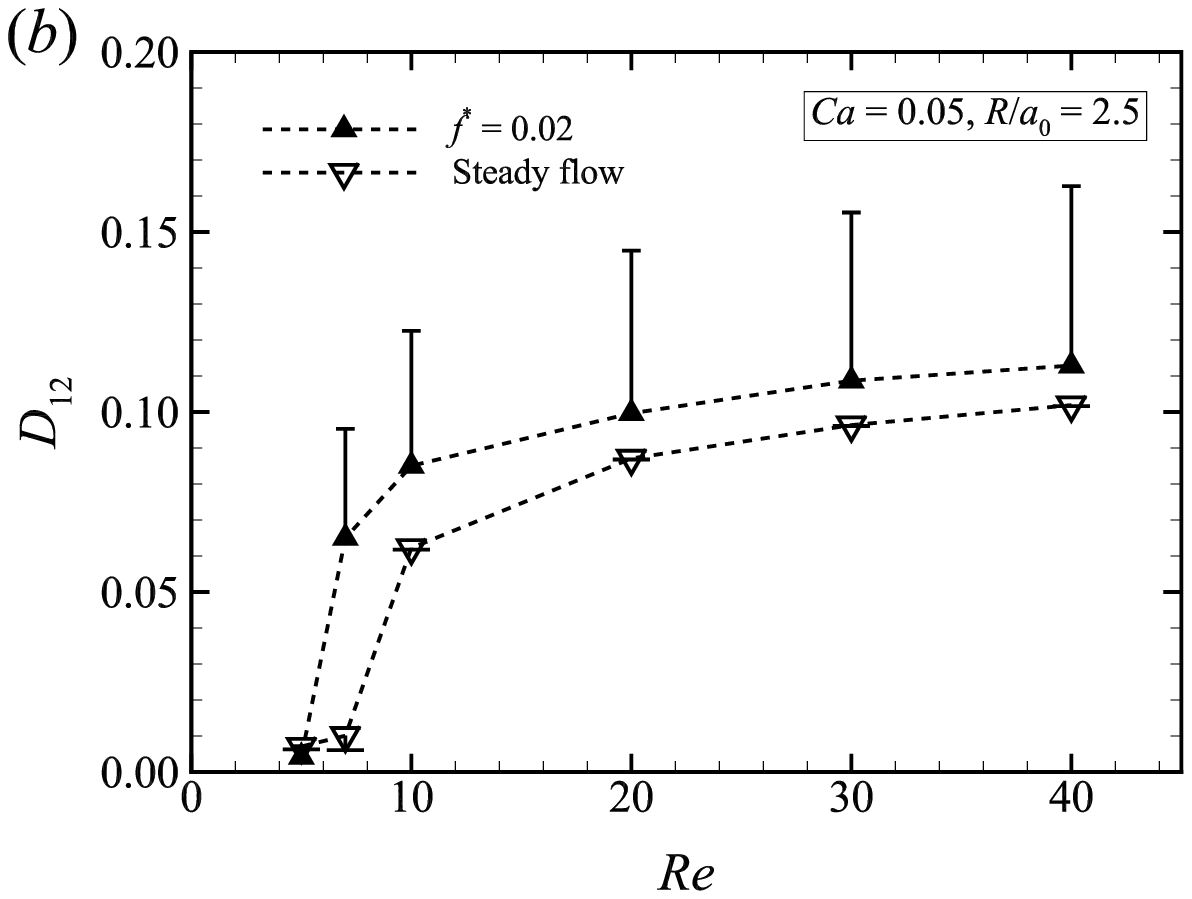}
  \includegraphics[height=4.5cm]{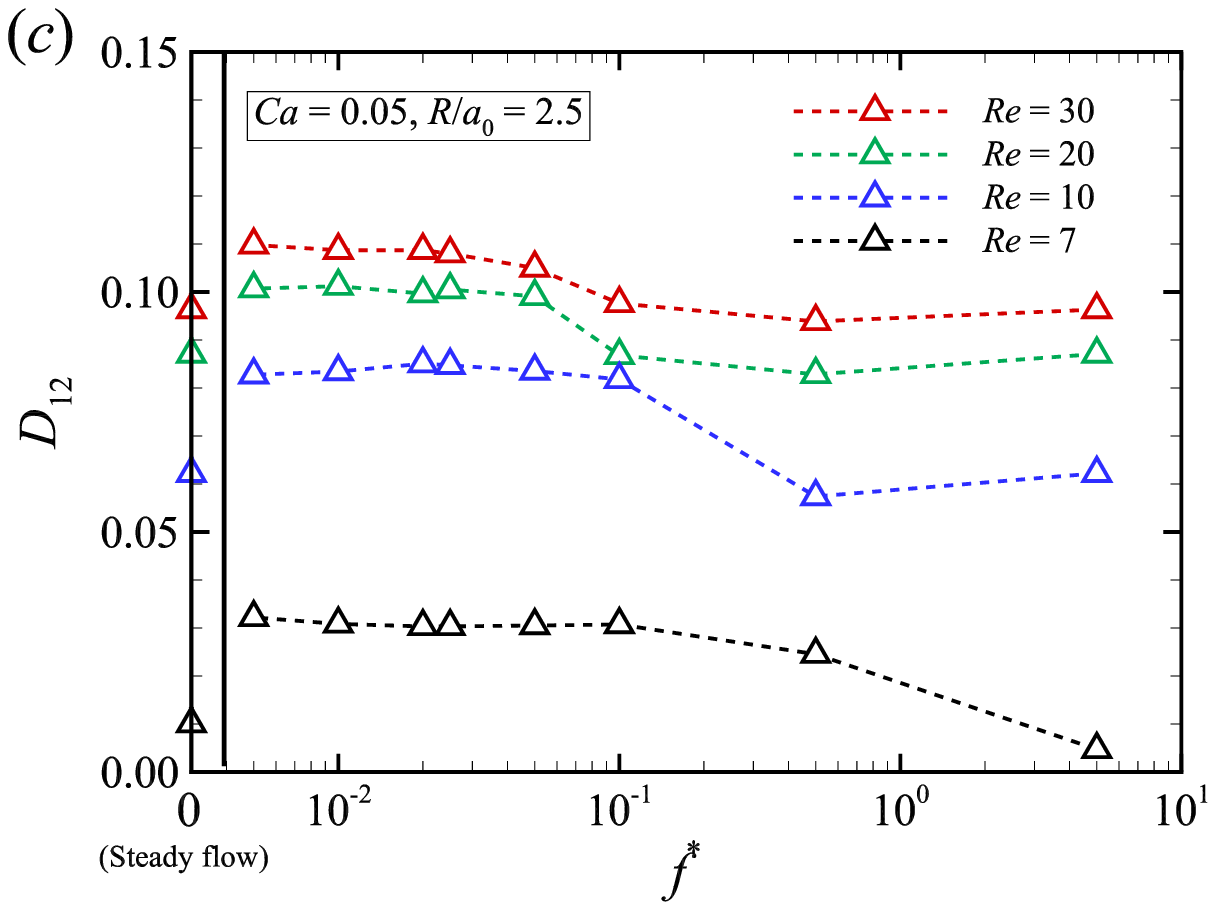}
  \includegraphics[height=4.5cm]{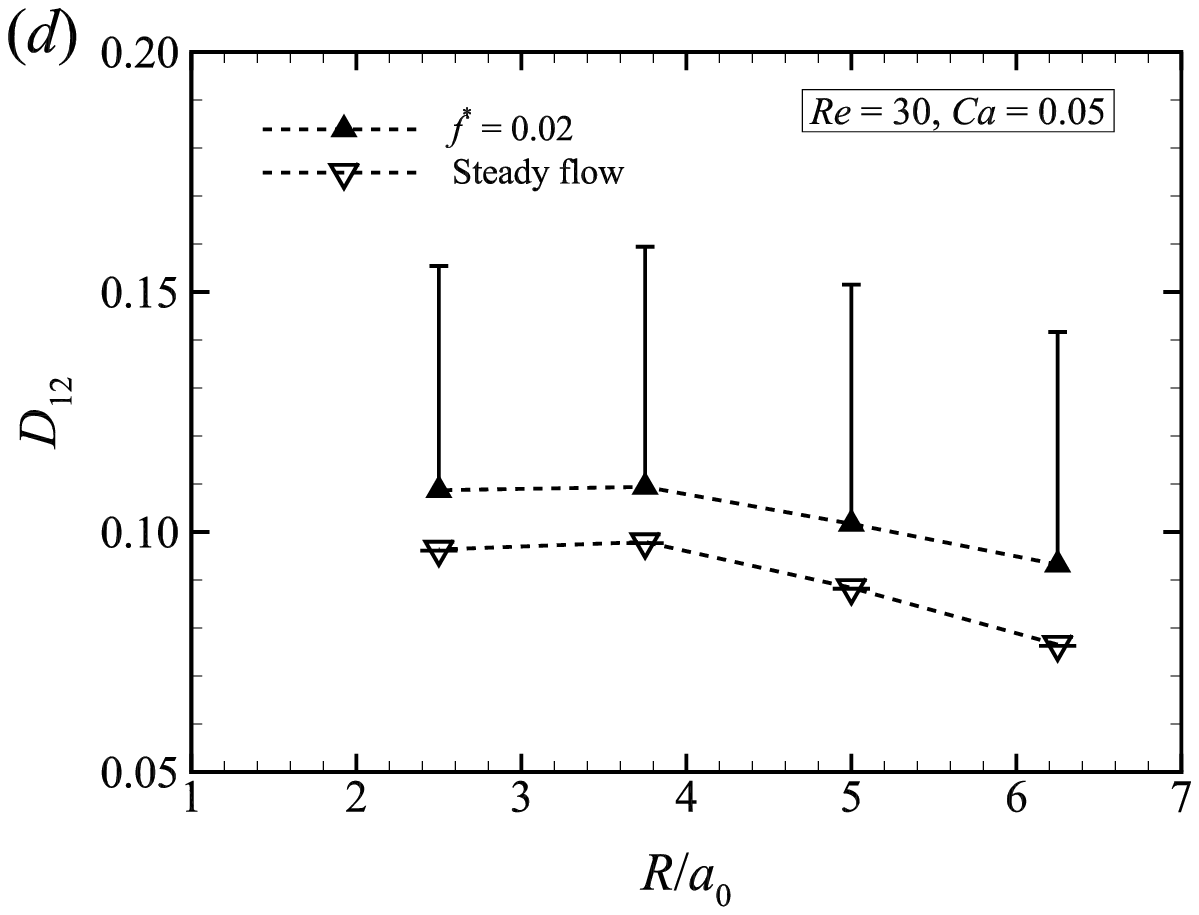}
  \caption{
         Time average of $\langle D_{12} \rangle$ as a function of ($a$) $Re$ (obtained with $Ca = 0.05$ and $R/a_0 = 2.5$), ($b$) $f^\ast$ (obtained with $Ca = 0.05$ and $R/a_0 = 2.5$), and
         ($c$) $R/a_0$ (obtained with $Re = 10$ and $Ca = 0.05$).
         The error bars in panels ($a$) and ($c$) are displayed only on one side of the mean value,
         and are not displayed in panel ($b$) for major clarity. 
  }
  \label{fig:D12}
\end{figure}

Figures~\ref{fig:D12}($a$--$c$) are the time average of $D_{12}$.
Overall, these results exhibit trends comparable to those of $\langle T_\mathrm{iso} \rangle$, previously shown in figures~\ref{fig:effect_re}($b$), \ref{fig:effect_f}($b$), and \ref{fig:effect_D}($c$).
Despite the similarities,
the axial-symmetric shaped capsule,
typical of large $Ca$,
exhibits small $D_{12}$ (figure~\ref{fig:D12}$a$),
and the capsule membrane state in pipe flows cannot be easily estimated from the deformed shape.
This is why we use the isotropic tension $T_\mathrm{iso}$ as an indicator of membrane deformation.

\section{\label{appA_vprof}Oscillatory velocity profile in a rigid tube}
Let us consider the Poiseuille flow in a rigid tube, with the radius of $R$, subject to the action of an oscillation pressure gradient,
as described by~\cite{Uchida1956, Womersley1955}.
The governing equation for oscillatory flow in cylindrical coordinates $(r, \theta, z)$ is
\begin{equation}
  \frac{\partial^2 v_z}{\partial r^2} + \frac{1}{r}\frac{\partial v_z}{\partial r} - \frac{1}{\nu}\frac{\partial v_z}{\partial t} = \frac{1}{\mu}\frac{\partial p}{\partial z},
  \label{govern_eq}
\end{equation}
where the pressure gradient can be represented by a Fourier series
\begin{equation}
  \frac{\partial p}{\partial z} = \sum_{k = 0}^\infty c_k e^{i k \omega t},
\end{equation}
with $c_0$ corresponding to the time average pressure gradient producing the Poiseuille profile.
The solution is sought in terms of the Fourier series
\begin{equation}
  v_z (r, t) = \sum_{k = 0}^\infty \hat{w}_k e^{i k \omega t}.
  \label{temp_solution}
\end{equation}
Inserting equation~\eqref{temp_solution} in equation~\eqref{govern_eq},
one gets
\begin{equation}
  \frac{d^2 \hat{w}_k}{d r^2} + \frac{1}{r}\frac{d \hat{w}_k}{d r} - \frac{i \omega k}{\nu} \hat{w}_k = \frac{c_k}{\mu},
  \label{temp_solution_2}
\end{equation}
where $i^2 = -1$, and defining the dimensionless variable $\zeta = r/R$,
the nonhomogeneous equation~\eqref{temp_solution_2} becomes
\begin{equation}
  \frac{d^2 \hat{w}_k}{d \zeta^2} + \frac{1}{\zeta}\frac{d \hat{w}_k}{d \zeta} - i k \alpha^2 \hat{w}_k = \frac{c_k}{\mu},
  \label{temp_solution_4}
\end{equation}
where $\alpha$ is the Womersley number,
\begin{equation}
  \alpha = R\sqrt{\frac{\omega}{\nu}}.
\end{equation}
With the boundary conditions
\begin{equation}
  \left. \hat{w}_k \right|_{r = R} = 0, \quad
  \left. \frac{d \hat{w}_k}{d r} \right|_{r = 0} = 0,
  \label{boundary}
\end{equation}
the particular solution of equation~\eqref{temp_solution_4} is easily found to be $-\frac{i \omega k}{\nu} \hat{w}_k = \frac{c_k}{\mu}$, and thus the global solution of the equation (for $k > 0$) is given by
\begin{equation}
  \hat{w}_k = \frac{ic_k R^2}{\mu k \alpha^2}
  + C_1 J_0 \left( \alpha \sqrt{k} \zeta i^{3/2} \right)
  + C_2 Y_0 \left( \alpha \sqrt{k} \zeta i^{3/2} \right),
  \label{temp_solution_5}
\end{equation}
where $C_1, C_2$ are arbitrary constants, and $J_0, Y_0$ are Bessel functions of order zero of the first and second kind, respectively.
Recall that $i^{1/2} = e^{i\pi/4} = (1 + i)/\sqrt{2}$.

The boundary conditions that the global solution must satisfy in a tube are the no-slip at the wall and the finite velocity along the axis of the tube, i.e.,
\begin{equation}
  \hat{w}_k (R) = 0, \quad
  |\hat{w}_k (0)| < \infty,
  \label{boundary_2}
\end{equation}
that provide the required conditions to determine the constants $C_1, C_2$.
It is known from the properties of $Y_0 (\zeta)$ that $Y_0 \to \infty$ as $\zeta$ (or $r$) goes to $0$.
Thus, the second boundary condition in equation~\eqref{boundary_2} leads to $C_2 = 0$,
and the first boundary condition then gives
\begin{equation}
  C_1 = \frac{- i c_k R^2}{\mu \alpha_k^2 J_0 (\alpha_k i^{3/2})},
\end{equation}
where $\alpha_k = \alpha\sqrt{k} = R\sqrt{k\omega/\nu}$.
With these values of $C_1, C_2$,
the solution $\hat{w}_k$ is finally
\begin{equation}
  \hat{w}_k = \frac{i c_k R^2}{\mu \alpha_k^2} \left( 1 - \frac{J_0 (\alpha_k \frac{r}{R} i^{3/2})}{J_0 (\alpha_k i^{3/2})} \right),
\end{equation}
and the velocity profile $v_z (r, t)$ is therefore
\begin{equation}
  v_z (r, t) = - \frac{c_0 R^2}{4 \mu} \left( 1 - \left( \frac{r}{R} \right)^2 \right)
  + \frac{R^2}{\mu} \sum_{k = 1}^\infty \mathfrak{R} \left( \frac{ic_k}{\alpha_k^2} \left[ 1 - \frac{J_0 (\alpha_k \frac{r}{R} i^{3/2})}{J_0 (\alpha_k i^{3/2})} \right] e^{i k \omega t} \right),
  \label{final_solution}
\end{equation}
where $\mathfrak{R}$ means the real part of a complex expression.

\begin{figure}%figure 16
  \centering
  \includegraphics[height=5.5cm]{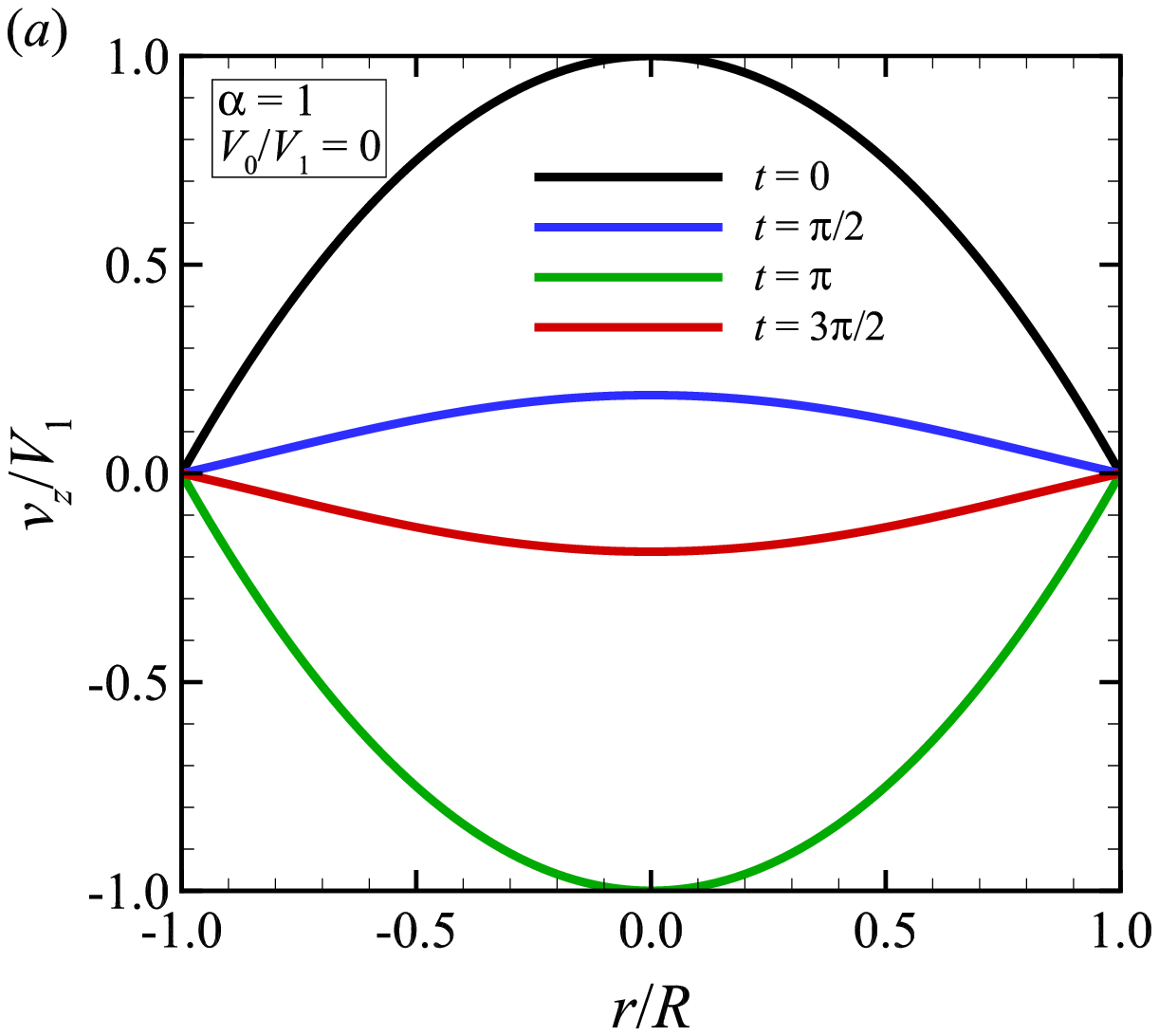}
  \includegraphics[height=5.5cm]{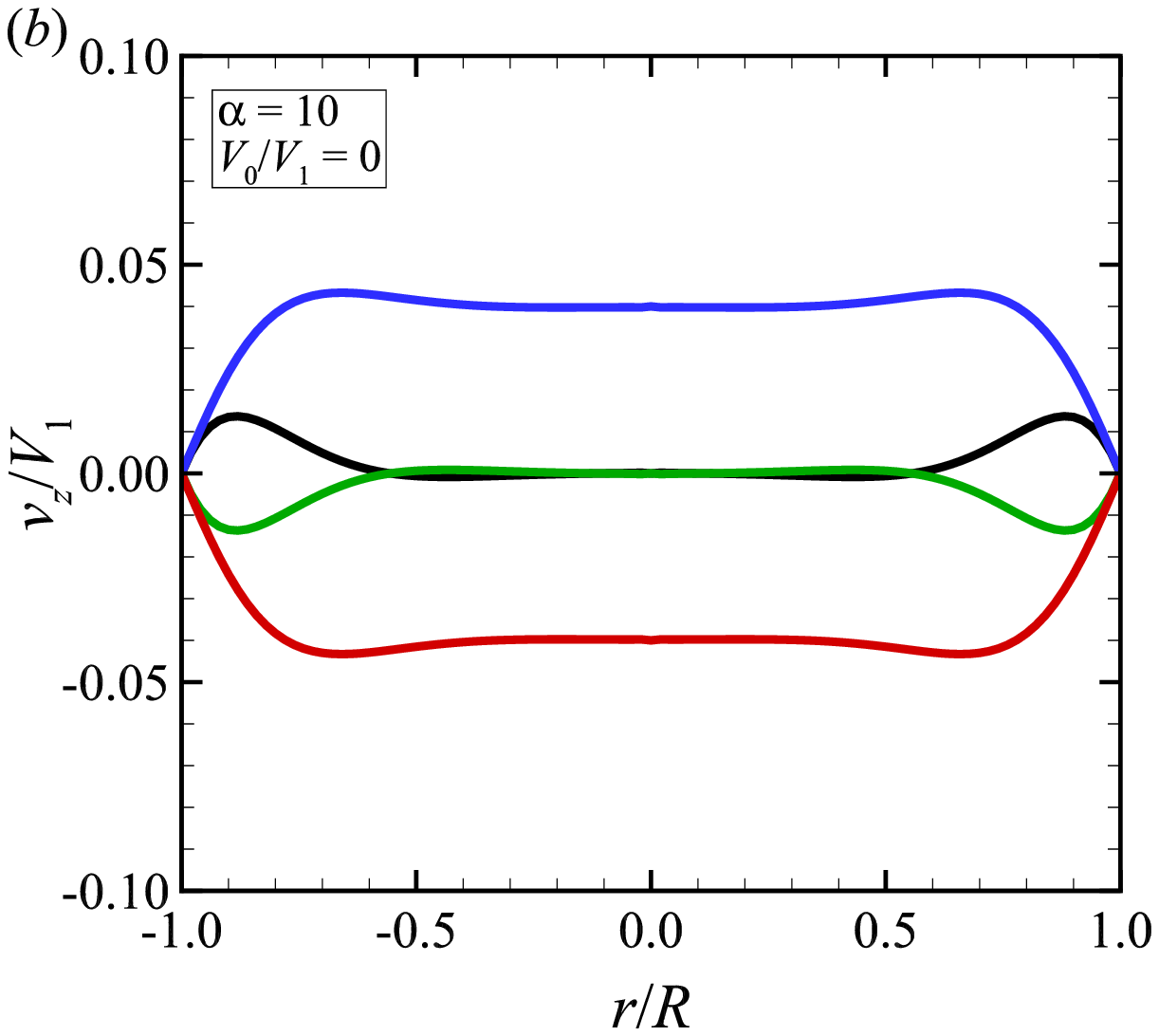}
  \includegraphics[height=5.5cm]{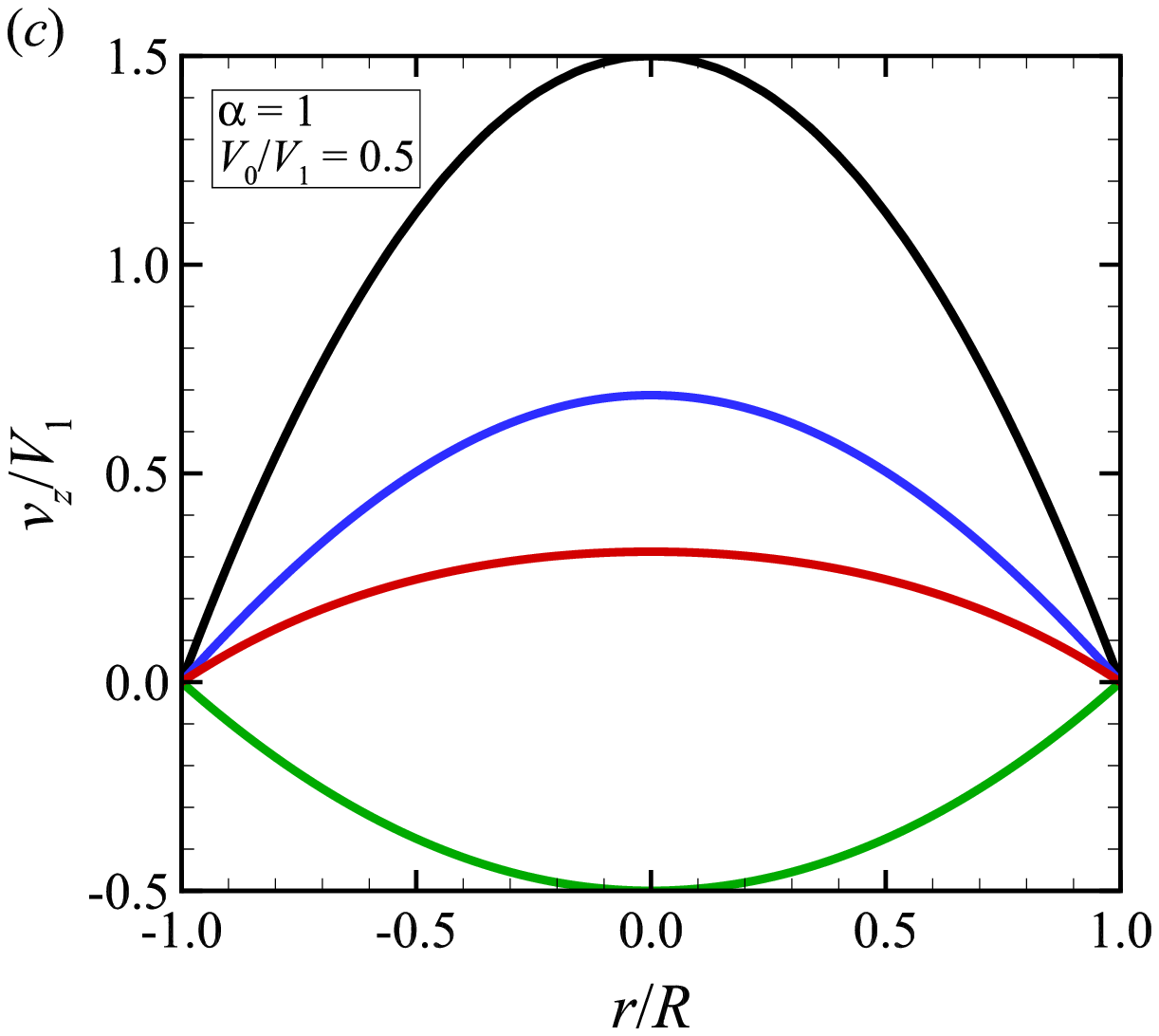}
  \includegraphics[height=5.5cm]{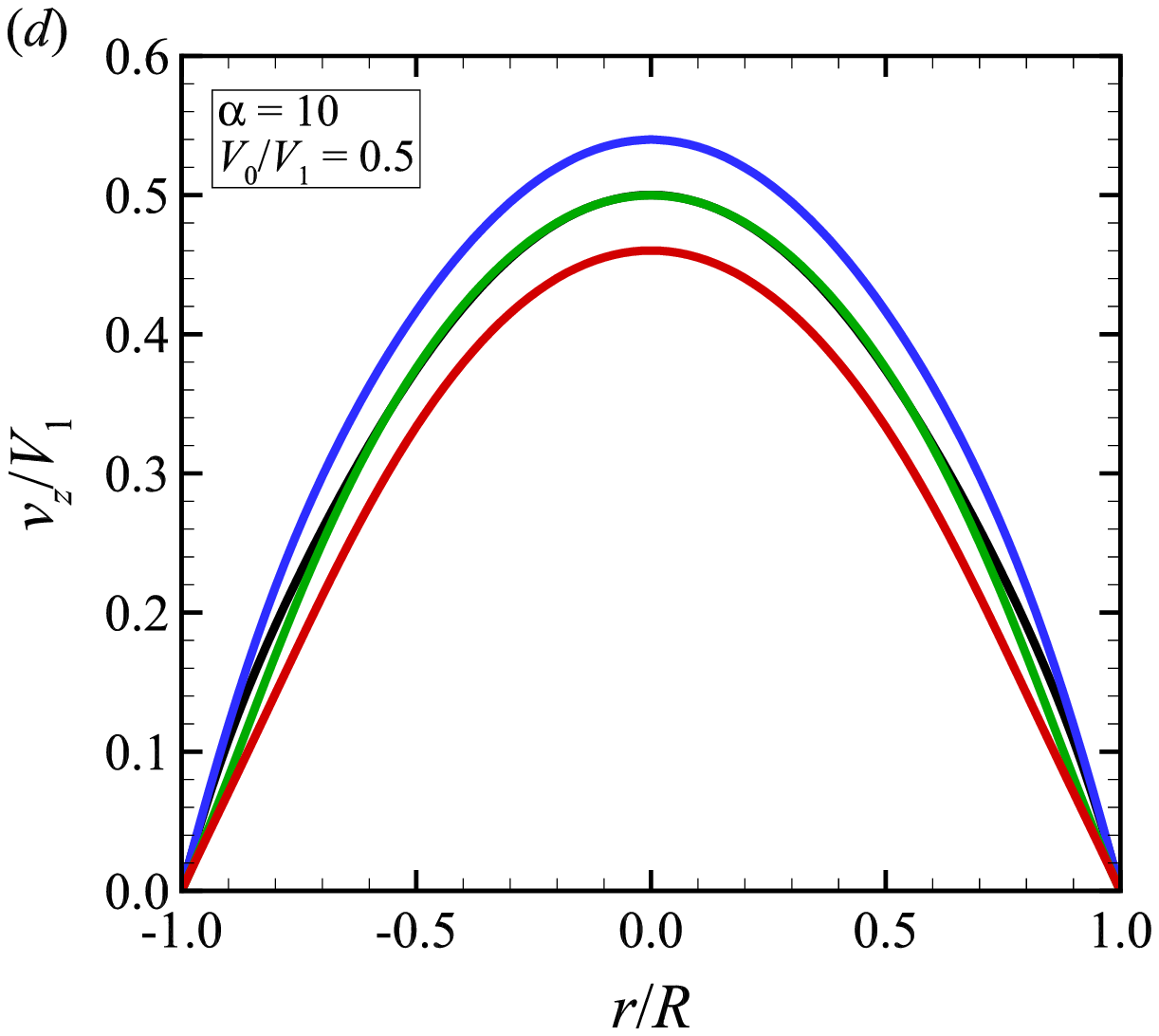}
  \caption{
         Oscillatory velocity profiles in a rigid tube at ($a$ and $c$) low frequency ($\alpha = 1$) and ($b$ and $d$) high frequency ($\alpha = 10$).
         The continuous Poiseuille component of the maximum velocity $V_0$ is neglected in ($a$) and ($b$),
         while the finite value of $V_0/V_1$ ($= 0.5$),
         which is the same condition used in the main text,
         is shown in ($c$) and ($d$).
         The lines represent the profiles at different phase angles ($\omega t$) within the oscillatory cycle, starting from $\omega t = 0$ and increasing by step of $\pi/2$.
  }
  \label{fig:vprofile}
\end{figure}

Let us consider oscillatory flow at low frequency, i.e., small $\alpha$.
The Taylor series of $J_0 (\zeta i^{3/2})$ is
\begin{equation}
  J_0 (\zeta i^{3/2}) 
  = 1 - \frac{(\zeta/2)^4}{(2!)^2} + \frac{(\zeta/2)^8}{(4!)^2} - \frac{(\zeta/2)^{12}}{(6!)^2} \cdots
  + i \left( \frac{(\zeta/2)^4}{(1!)^2} + \frac{(\zeta/2)^6}{(3!)^2} - \frac{(\zeta/2)^{10}}{(5!)^2} - \cdots \right),
\end{equation}
and taking the dominant terms into account,
one obtains
\begin{equation}
  \frac{J_0 (\alpha_k \frac{r}{R} i^{3/2})}{J_0 (\alpha_k i^{3/2})}
  = 1 - i \frac{\alpha_k^2}{4} \left( 1 - \left( \frac{r}{R} \right)^2 \right)
  - \frac{\alpha_k^4}{64} \left( 3 - 4 \left( \frac{r}{R} \right)^2 + \left( \frac{r}{R} \right)^4 \right) \sin{\omega t} + O(\alpha_k^6).
\end{equation}
The velocity is thus written as
\begin{align}
  v_z (r, t)
  &= - \frac{c_0 R^2}{4 \mu} \left( 1 - \left( \frac{r}{R} \right)^2 \right) \\
  &- \frac{R^2}{\mu} \sum_{k = 1}^\infty \left\{ \frac{c_k}{4} \left( 1 - \left( \frac{r}{R} \right)^2 \right) \cos{(k\omega t)} + \frac{c_k \alpha_k^2}{64} \left( 3 - 4 \left( \frac{r}{R} \right)^2 + \left( \frac{r}{R} \right)^4 \right) \sin{(k\omega t)} \right\},
\end{align}
%If the continuous Poiseuille component of the velocity is neglected and
and if we set $V_i = - \frac{c_i R^2}{4 \mu}$ ($i = 0, 1$) and $\alpha_1 = \alpha$, the first mode $k = 1$ becomes
\begin{equation}
  \frac{v_z}{V_1} =
  \frac{V_0}{V_1} \left( 1 - \left( \frac{r}{R} \right)^2 \right)
  + \left( 1 - \left( \frac{r}{R} \right)^2 \right) \cos{(\omega t)}
  + \frac{\alpha^2}{16} \left( 3 - 4 \left( \frac{r}{R} \right)^2 + \left( \frac{r}{R} \right)^4 \right) \sin{(\omega t)}.
\end{equation}
Figure~\ref{fig:vprofile}($a$) shows the velocity profile for $\alpha = 1$ at each phase angle $\omega t (= 0, \pi/2, \pi, 3\pi/4$) when the Poiseuille component of the velocity is neglected (i.e., $V_0/V_1 = 0$).
Starting from the Poseuille flow at time $\omega t = 0$, at the phase of $\omega t = \pi/2$,
the Poiseuille is still positive while the corresponding pressure gradient vanishes.
The phase difference disappears at mid cycle ($\omega t = \pi$) when the Poiseuille flow is recovered. The profile reaches its peak form at the peak pressure gradient ($\omega t = 0, \pi$).
The velocity profile for $V_0/V_1 = 0.5$,
which is the same condition discussed in the main text,
is also shown in figure~\ref{fig:vprofile}($c$) for completeness.

At high frequency, the oscillatory flow in a rigid tube is less able to keep pace with the changing pressure,
thus reaching less than the fully developed Poiseuille flow profile at the peak of each cycle.
The parameter $\alpha r/R$ takes large values and the axis ($r = 0$) is excluded from the analysis.
For high values of its argument,
the asymptotic development of $J_0 (\zeta)$ is such that
\begin{equation}
  J_0 (\zeta) = \sqrt{\frac{2}{\pi \zeta}} \cos{\left( \zeta - \frac{\pi}{4} \right)} + O(|\zeta|^{-1}),
  \quad \text{with} \quad |arg(\zeta)| < 2 \pi.
\end{equation}
Using the relation $i^{3/2} = e^{i 3\pi/4}$ and $s = \alpha_k r/R$,
we can perform the following algebraic calculation
\begin{align}
  J_0 (e^{i 3 \pi/4}s)
  &= e^{- 3 \pi/8} \sqrt{\frac{2}{\pi s}} \cos{\left( e^{i 3 \pi/4}s - \frac{\pi}{4} \right)} \\
  &= e^{- i 3 \pi/8} \sqrt{\frac{2}{\pi s}} \cosh{\left( \frac{s}{\sqrt{2}} + i \left( \frac{s}{\sqrt{2}} + \frac{\pi}{4}  \right) \right)},
\end{align}
and neglecting the decaying exponential in cosh, since we deal with large values of the argument, we obtain
\begin{equation}
  J_0 (e^{i 3 \pi/4}s) = e^{- 3 \pi/8} \frac{1}{\sqrt{2 \pi s}} e^{\frac{s}{\sqrt{2}}} e^{i \left( \frac{s}{\sqrt{2}} + \frac{\pi}{4}  \right)},
\end{equation}
which leads us to find
\begin{equation}
  \frac{J_0 (\alpha_k \frac{r}{R} i^{3/2})}{J_0 (\alpha_k i^{3/2})}
  \approx \frac{1}{\sqrt{r/R}} e^{-(1+i) \frac{\alpha_k}{\sqrt{2}} \left(1 - \frac{r}{R} \right)}.
\end{equation}
%If the continuous Poiseuille component of the velocity is neglected,
Finally, the first mode of the velocity profile yields
\begin{align}
  \frac{v_z}{V_1}
  &= \frac{V_0}{V_1} \left( 1 - \left( \frac{r}{R} \right)^2 \right) \nonumber \\
  &+ \frac{4}{\alpha^2} \left[ \sin{(\omega t)} - \frac{1}{\sqrt{r/R}} e^{-\frac{\alpha}{\sqrt{2}} \left( 1 - \frac{r}{R} \right)} \sin{\left( \omega t - \frac{\alpha}{\sqrt{2}} \left(1 - \frac{r}{R} \right) \right)} \right] 
  + O\left( \frac{1}{\alpha^4} \right).
\end{align}
Figure~\ref{fig:vprofile}($b$) shows the velocity profile for $\alpha = 10$ at each phase angle, when the continuous Poiseuille component of the velocity is neglected.
While the velocity is everywhere close to zero,
the profile reaches its peak form at the phase of $\omega t = \pi/2$ (and $3\pi/2$),
i.e., the resulting flow is in complete phase shift (by $\pi/2$) with respect to the pressure gradient.
The velocity profile for $V_0/V_1 = 0.5$ is also shown in figure~\ref{fig:vprofile}($d$) for completeness.
Thus, in this case the continuous Poiseuille component is retained at each time.

\bibliographystyle{jfm}
% Note the spaces between the initials
\bibliography{jfm-instructions}

\end{document}